\documentclass[prl,twocolumn,superscriptaddress,preprintnumbers]{revtex4-1}
\usepackage{graphicx}% Include figure files
\usepackage{dcolumn}% Align table columns on decimal point
\usepackage{bm}% bold math
\usepackage{subfigure}
\usepackage{wrapfig}
\usepackage{amssymb,latexsym, amsmath}
\usepackage{graphics}
\usepackage{caption}
\usepackage{float}
\usepackage{appendix}

\begin{document}

% Use the \preprint command to place your local institutional report
% number in the upper righthand corner of the title page in preprint mode.
% Multiple \preprint commands are allowed.
% Use the 'preprintnumbers' class option to override journal defaults
% to display numbers if necessary
%\preprint{}

%Title of paper
\title{Stripe-like nanoscale structural phase separation and optimal inhomogeneity in superconducting BaPb$_{1-x}$Bi$_x$O$_3$}

% repeat the \author .. \affiliation  etc. as needed
% \email, \thanks, \homepage, \altaffiliation all apply to the current
% author. Explanatory text should go in the []'s, actual e-mail
% address or url should go in the {}'s for \email and \homepage.
% Please use the appropriate macro foreach each type of information

% \affiliation command applies to all authors since the last
% \affiliation command. The \affiliation command should follow the
% other information
% \affiliation can be followed by \email, \homepage, \thanks as well.
\author{P. \surname{Giraldo-Gallo}}
%\email[]{pgiraldo@stanford.edu}
%\homepage[]{Your web page}
%\thanks{}
%\altaffiliation{}
\affiliation{Geballe Laboratory for Advanced Materials, Stanford University, Stanford, CA 94305, USA}
\affiliation{Department of Physics, Stanford University, CA 94305, USA}
\author{Y. Zhang}
%\author{M. J. Kramer}
\affiliation{Ames Laboratory (USDOE), Department of Materials Science and Engineering, Iowa State University, Ames IA 50011-3020, USA}
\affiliation{Beijing National Laboratory for Condensed Matter Physics, Institute of Physics, Chinese Academy of Sciences, Beijing 100190, China}
%\affiliation{Materials and Engineering Physics, Ames Laboratory (USDOE), Iowa State University, Ames, IA 50011-3020, USA}
%\affiliation{Department of Materials Science and Engineering, Iowa State University, Ames, IA 50011-3020, USA}
\author{C. Parra}
\affiliation{Geballe Laboratory for Advanced Materials, Stanford University, Stanford, CA 94305, USA}
\affiliation{Department of Physics, Stanford University, CA 94305, USA}
\affiliation{Stanford Institute for Materials and Energy Sciences, SLAC National Accelerator Laboratory, Menlo Park, CA 94025, USA}
\affiliation{Departmento de F\'isica, Universidad T\'ecnica Federico Santa Mar\'ia, Valpara\'iso, Chile}
\author{H. C. Manoharan}
\affiliation{Geballe Laboratory for Advanced Materials, Stanford University, Stanford, CA 94305, USA}
\affiliation{Department of Physics, Stanford University, CA 94305, USA}
\affiliation{Stanford Institute for Materials and Energy Sciences, SLAC National Accelerator Laboratory, Menlo Park, CA 94025, USA}
\author{M. R. Beasley}
\author{T. H. Geballe}
%\email[]{geballe@stanford.edu}
\affiliation{Geballe Laboratory for Advanced Materials, Stanford University, Stanford, CA 94305, USA}
\affiliation{Department of Applied Physics, Stanford University, CA 94305, USA}
\author{M. J. Kramer}
\affiliation{Ames Laboratory (USDOE), Department of Materials Science and Engineering, Iowa State University, Ames IA 50011-3020, USA}
\author{I. R. Fisher}
%\email[]{irfisher@stanford.edu}
\affiliation{Geballe Laboratory for Advanced Materials, Stanford University, Stanford, CA 94305, USA}
\affiliation{Department of Applied Physics, Stanford University, CA 94305, USA}
%\affiliation{Geballe Laboratory for Advanced Materials and Department of Applied Physics, Stanford University, Stanford, California 94305, USA \\}
%\affiliation{Applied Physics Department, Stanford University, California, 94305, USA \\}

%Collaboration name if desired (requires use of superscriptaddress
%option in \documentclass). \noaffiliation is required (may also be
%used with the \author command).
%\collaboration can be followed by \email, \homepage, \thanks as well.
%\collaboration{}
%\noaffiliation

\date{\today}

\begin{abstract}
Structural phase separation in the form of partially disordered stripes, with characteristic length scales in the nanometer range, is observed for superconducting BaPb$_{1-x}$Bi$_x$O$_3$. The evolution of the superconducting coherence length with composition relative to the size of these stripes suggests an important role of the nanostructure in determining the shape of the superconducting dome. It is proposed that the maximum $T_c$ is determined by a kind of ``optimal inhomogeneity'', characterized by a crossover from an inhomogeneous macroscopic superconductor to a granular superconductor for which phase fluctuations suppress $T_c$. 
\end{abstract}

% insert suggested PACS numbers in braces on next line
\pacs{74.81.-g,74.62.En,74.40.Kb}
% insert suggested keywords - APS authors don't need to do this
%\keywords{}

%\maketitle must follow title, authors, abstract, \pacs, and \keywords
\maketitle

%\section{Introduction}

High temperature superconductors (HTSCs) are complex materials with many degrees of freedom, including spin, charge, orbital and structural. Spontaneous segregation of electronic phases potentially plays an important role in defining several important physical properties in these materials, including the critical temperature $T_c$. 
%Such phenomena are not exclusive to superconducting systems, and are found in many other strongly correlated systems \cite{Dagotto1, Dagotto2, Akahoshi}, stimulating considerable experimental and theoretical effort to understand their causes and consequences. 
Phase separation in the charge channel has been observed in underdoped cuprates in the form of stripes \cite{Tranquada1, Howald2}, and recently, in the form of charge density wave (CDW) nano-domains \cite{Letacon}. Details of the nanostructure associated with this phase separation, and its connection with the optimization of $T_c$ in these systems is an important open question \cite{Kivelson4, Kivelson5, kivelson6}. For systems with such a variety of interactions, tracking the influence of each individual degree of freedom on the phase separation and on the determination of the electronic properties is challenging. For this reason, the study of simpler superconducting systems can provide useful insights for understanding more complex materials. A model system for the study of how superconductivity is influenced by local CDW instabilities and structural phase separation can be found in the bismuthate superconductors.

\begin{figure}[!htbp]
\vspace{-0.5cm}
\hspace{-0.6cm}
\centering
\includegraphics[scale=0.32]{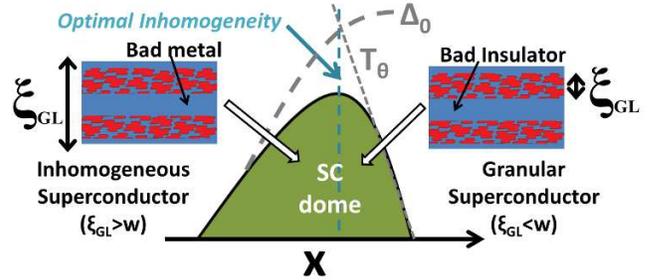}\\% \vspace{-0.3cm}%\hspace{-0.9cm}\\
\vspace{-0.6cm}
\caption{(Color online) Stylized cartoon illustrating the effect of the competing length scales associated with structural phase separation and the superconducting coherence length on the superconducting phase diagram of BaPb$_{1-x}$Bi$_x$O$_3$ (see discussion in main text). Red and blue regions correspond to tetragonal and orthorhombic polymorphs respectively. The pairing interaction is understood to originate in the tetragonal material \cite{Cava1}.
%For compositions below optimal doping (left side of diagram), the material forms partially disordered stripes comprising metallic/non-superconducting orthorhombic phase (blue) and metallic/superconducting tetragonal phase (red). In this regime, the characteristic size of the stripes is smaller than $\xi_{GL}$. The material is a bulk inhomogeneous superconductor, and $T_c$ is determined by the average pairing energy scale, $\bar{\Delta}_0$. Beyond optimal doping (right side of diagram),  the stripe formation comprises a more strongly insulating orthorhombic phase (blue) and a metallic/superconducting tetragonal phase (red). The characteristic length scale of the stripes in this regime is comparable to $\xi_{GL}$, and the material behaves as a granular superconductor, with the bulk $T_c$ suppressed by phase fluctuations. (See discussion in main text.)
}\label{fig_BPBOmodel}
%\vspace{-0.4cm}
\end{figure}

\begin{figure}[ht]
%\vspace{-0.5cm}
\hspace{1cm}
\centering
%\hspace{-1.03cm}
\hspace{1cm}
\includegraphics[scale=0.4]{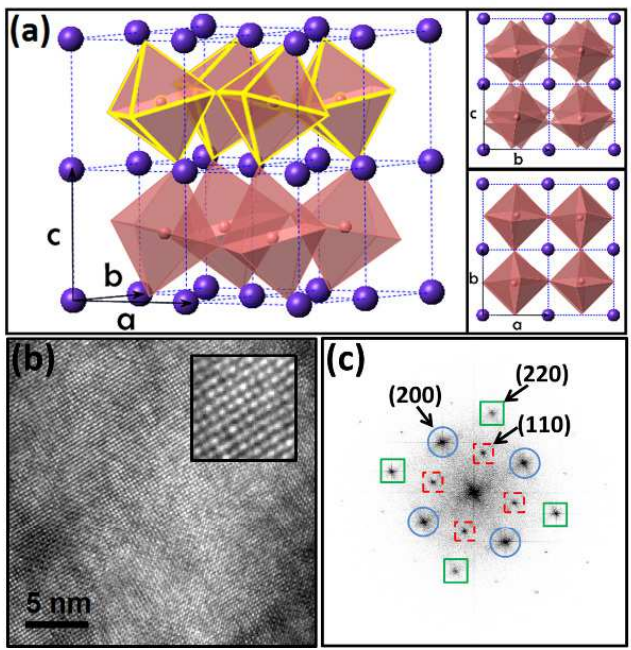} \\ %\vspace{-0.1cm}\hspace{-0.6cm}
\vspace{-0.2cm}
\caption{(Color online)  {\bf (a)} Schematic diagram illustrating the $a^0b^-b^-$ distortion of the cubic perovskite structure, resulting in a orthorhombic \textit{Ibmm} unit cell. 
%The main panel shows a perspective view, and smaller panels show projections along the a and c axes (for simplicity, axes are labeled relative to the original cubic cell). 
The rotations of the octahedra have been exaggerated in the illustrations to more clearly reveal the distortion. 
%For the real material, typical rotations are between 6.7$^{\circ}$ and 8.7$^{\circ}$ depending on the composition \cite{Cava1}. 
{\bf (b)} HRTEM image for a sample with $x$=0.18, looking down the [001] zone axis, revealing a well-ordered structure with coherent planes of atoms. Inset shows an expanded view of a region of 3$\times$3nm$^2$. {\bf (c)} FFT of the image in fig. \ref{fig_crystal}(b), showing peaks corresponding to the $[200]$ (blue-solid circles), $[110]$ (red-dashed squares) and $[220]$ (green-solid squares) set of reflections. Symbols with solid edges correspond to reflections allowed in both \textit{Ibmm} and \textit{I4/mcm} space groups, while those with dashed edges correspond to reflections forbidden in the \textit{I4/mcm} space group, but allowed in the \textit{Ibmm}.}\label{fig_crystal}
\vspace{-0.4cm}
\end{figure}

The family of bismuthate superconductors results from replacing K for Ba, or Pb for Bi, in BaBiO$_3$, a charge density wave (CDW) insulator \cite{Uchida1, Tarapheder1, Gabovich1, Baumert1}. This family of superconductors has no magnetic degrees of freedom. Upon doping, the insulating CDW phase disappears, giving rise to a metallic phase where superconductivity appears at (maximum) temperatures below 30K and 11K for K-doping and Pb-doping, respectively \cite{Cava2, Sleight1}. Structural phase separation on a nanoscopic scale has been observed in BaPb$_{1-x}$Bi$_x$O$_3$ for superconducting compositions \cite{Paula1}, but the implications in shaping the superconducting dome, in particular close to the disorder induced metal-insulator transition at $x=$ 0.30 \cite{Katherine1} followed closely by the opening of a gap in the optical spectrum at $x\approx$ 0.35, remain an open question. %Remarkably, analysis of recent point contact spectroscopy measurements \cite{Katherine1} suggest that this material is intrinsically a 30K-range superconductor, just like the other family member of the bismuthate superconductors, Ba$_{1-x}$K$_x$BiO$_3$, but where $T_c$ has been suppressed by disorder of some sort. 
In this letter we report the observation of stripe-like structural phase separation in superconducting BaPb$_{1-x}$Bi$_x$O$_3$ for compositions spanning optimal doping. Remarkably, the maximum $T_c$ occurs when the superconducting coherence length matches the size of the partially disordered stripes, implying a connection between the structural phase separation, the enhanced coulomb effects due to disorder (localization), the inhomogeneous superconducting properties, and the shape of the superconducting ``dome" (see figure \ref{fig_BPBOmodel}).

%\begin{figure}[ht]
%\centering
%\hspace{-0.7cm}
%\includegraphics[scale=0.35]{2825_2_440Kx_abs_0andby16resol_withoutGr_invertedScale.eps} \hspace{-0.6cm}
%\vspace{-0.3cm}
%\caption{(Color online) {\bf (a)} 19$\times$19$nm^2$ portion of a typical $[101]$ filtered-and-reconstructed HRTEM image for a sample with bismuth concentration of $x$=0.18, showing the atomic %resolution detail. {\bf (b)} Same image as in (a) after a 7.5$\AA\times$7.5$\AA$ averaging, eliminating the atomic resolution information while maintaining the broader orthorhombic structural variation. Vertical bars show intensity color scale. Blue regions are more strongly orthorhombic, red regions are less strongly orthorhombic (i.e. more strongly tetragonal).}\label{fig_redresol}
%\vspace{-0.4cm}
%\end{figure}

BaPb$_{1-x}$Bi$_x$O$_3$ has a distorted perovskite (ABO$_3$) crystal structure. %Distortions from the ideal structure, which is cubic with a single `B' site, fall in to two distinct types. First, 
For the highest Bi concentrations the material comprises two distinct Bi sites, with different Bi-O bond lengths. The origin of the associated charge density wave (CDW) has been widely debated \cite{Mattheiss1,Varma1,Tarapheder1,Yin1}. For $x \leq$ 0.8 the average structure comprises a single Bi/Pb site \cite{Marx1}, though EXAFS measurements reveal two distinct Bi-O bond lengths down to at least $x \sim $ 0.25 \cite{Boyce1}, implying a persistence of the CDW at a local level. Significantly, for all compositions, the perovskite structure is also distorted by rotational instabilities of the oxygen octahedra, which can be described using Glazer's notation \cite{Glazer, Howard}. %The tilt axis varies for different Bi concentrations, resulting in different unit cell symmetries across the phase diagram. 
For the insulating end-member compound BaBiO$_3$ ($x=1$), and down to $x=0.9$, the unit cell space group is monoclinic \textit{I2/m} (coming from a $a^0b^-c^-$ tilt, in Glazer's notation); for the metallic end-member compound BaPbO$_3$ ($x=0$) and up to $x\approx 0.15$, and again for $0.35<x<0.9$, the unit cell space group is orthorhombic \textit{Ibmm} (coming from a $a^0b^-b^-$ tilt, as shown in fig. \ref{fig_crystal}(a)); however, for the region of $0.15<x<0.35$, which is also the range of compositions for which the material is superconducting, the material is polymorphic, with a fraction of its volume with orthorhombic \textit{Ibmm} symmetry and the rest with tetragonal \textit{I4/mcm} symmetry (coming from a $a^0a^0c^-$ tilt) \cite{Cava1}. The superconducting volume fraction peaks at the same Bi composition where the tetragonal-to-orthorhombic ratio is maximum, leading to the conclusion that the tetragonal polymorph is the one responsible for superconductivity in this material \cite{Cava1,Marx1}. This Bi composition is also the one for which the material has the maximum $T_c$, i.e., the optimal doping composition.

%In order to investigate 
In the interest of investigating how the polymorphism is accommodated microscopically in a ``single crystal'' of BaPb$_{1-x}$Bi$_x$O$_3$, and its possible consequences for the observed transport and, more interestingly, superconducting properties, high-resolution transmission electron microscopy (HRTEM) measurements were taken for samples with bismuth compositions below, at and above optimal doping. Samples of each concentration were crushed in liquid-nitrogen-cooled ethyl alcohol, and the liquid was allowed to warm to room temperature. The slurry was stirred and a small droplet was placed on a holey carbon grid and dried in air. The samples were analyzed using a FEI G2 F20TEM Tecnai STEM operated at 200 keV. Thin areas were analyzed with selected area diffraction, energy dispersive spectroscopy, and high-resolution imaging. Thin areas were aligned with either the $[$010$]$ or the $[$001$]$ zone axis based on indexing to the \textit{Ibmm} structure (space group No. 74), showing clear lattice fringes in the HRTEM. All the HRTEM images taken for all the different compositions reveal a well-ordered structure, as can be observed in the 24.1$\times$24.1nm$^2$ image in fig. \ref{fig_crystal}(b) for a sample with Bi composition of $x=0.18$, and better appreciated in the 3$\times$3nm$^2$ expanded view in the inset to this figure. Fig. \ref{fig_crystal}(c) shows its corresponding fast Fourier transform (FFT), revealing peaks from both tetragonal (hkl even) and orthorhombic (hkl even and odd, in the tetragonal notation) phases. In ref. \cite{Paula1} we showed that it is possible to recreate the spatial separation of the two polymorphs by systematically masking these diffraction peaks and performing an inverse Fourier transform (IFFT). In this way, we can obtain information about the length scales associated with the polymorphic variation across a sample.

%Fig. \ref{fig_redresol}(a) shows a 19$\times$19nm$^2$ portion of the $[$101$]_T$ filtered IFFT of the HRTEM image in fig. \ref{fig_crystal}(b). This image keeps the information of both the atomic periodicity as well as a larger-scale contrast variation, reflecting variations in the local ``orthorhombicity'' across the sample. The image in figure \ref{fig_redresol}(b) is the result of a resolution reduction by adjacent averaging, of the image in fig. \ref{fig_redresol}(a), from 0.47$\AA$ per pixel, to 7.5$\AA$ per pixel, therefore eliminating the atomic resolution information while keeping the longer-range variation in ``orthorhombicity''. For the purpose of the following analysis, we will consider only the reduced resolution images, given that these conserve the information of the longer-scale structural variation while reducing the computational requirements (see supplemental material). 

\begin{figure}[ht]
%\vspace{-0.7cm}
\centering
%\hspace{-1.03cm}
\hspace{-1cm}
\includegraphics[scale=0.68]{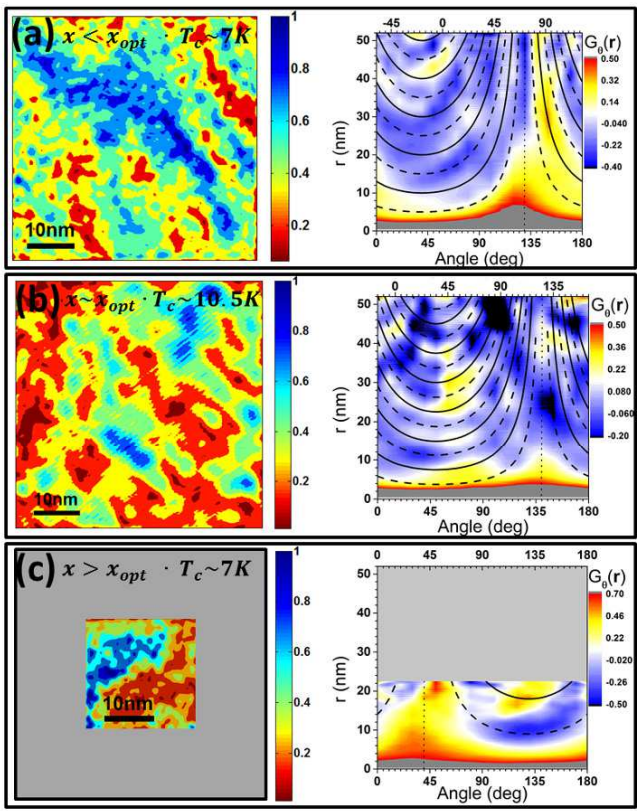} 
\hspace{-0.9cm}
\vspace{-0.2cm}
%\caption{(Color online) $[101]_T$ filtered-and-reconstructed HRTEM images (after averaging-out the atomic-scale variation), and their respective correlation functions for bismuth compositions of {\bf (a)} $x$=0.18 (i.e. $x<x_{opt}$, $T_c$ $\approx$ 7K) {\bf (b)} $x$=0.24 ($x \sim x_{opt}$, $T_c$ $\approx$ 10.5 K), and {\bf (c)} $x$=0.28 ($x>x_{opt}$, $T_c$ $\approx$ 7 K). For each horizontal panel, the center column shows the corresponding average spatial correlation function, $\langle G(\vec{r})\rangle$. Solid blue vertical lines indicate the local maxima in $\langle G(\vec{r})\rangle$, while dashed-black vertical lines indicate local minima. The third column in each horizontal panel shows the angle-dependent spatial correlation function, $\langle G_{\theta}(\vec{r})\rangle$ on a color scale, as a function of $|\vec{r}|$ (vertical axis) and the angle $\theta$ with the horizontal (bottom-axis) or the $\left[200\right]_T$ crystalline axis (top-axis). Solid and dashed lines represent the best fits to $N\times d/\cos((\alpha-90^{\circ})-\theta)$ and $(2N-1)\times w/\cos((\alpha-90^{\circ})-\theta)$ for the local maxima and minima respectively, as described in the main text.}\label{fig_HRTEM}
\caption{(Color online) $[110]_T$/$[101]_T$ filtered-and-reconstructed HRTEM images (after averaging-out the atomic-scale variation), and their respective angle-dependent correlation functions, $\langle G_{\theta}(\vec{r})\rangle$, for bismuth compositions of {\bf (a)} $x$=0.18 (i.e. $x<x_{opt}$, $T_c$ $\approx$ 7K) {\bf (b)} $x$=0.24 ($x \sim x_{opt}$, $T_c$ $\approx$ 10.5 K), and {\bf (c)} $x$=0.28 ($x>x_{opt}$, $T_c$ $\approx$ 7 K). Vertical bars next to each image show the normalized intensity color scale. Blue regions are more strongly orthorhombic, red regions are less strongly orthorhombic (i.e. more strongly tetragonal). The color scale on the right hand side plots represent the value of $\langle G_{\theta}(\vec{r})\rangle$. This quantity is plotted as a function of $|\vec{r}|$ (vertical axis) and the angle $\theta$ with the horizontal (bottom-axis) or the $\left[200\right]_T$ crystalline axis (top-axis). Solid and dashed lines represent the best fits to $N\times d/\cos((\alpha-90^{\circ})-\theta)$ and $(2N-1)\times w/\cos((\alpha-90^{\circ})-\theta)$ for the local maxima and minima respectively, as described in the main text.}\label{fig_HRTEM}
\vspace{-0.4cm}
\end{figure}

%Fig. \ref{fig_HRTEM} shows reduced resolution filtered-and-reconstructed HRTEM images (left panels) for a representative sample of each Bi composition studied, together with the average spatial correlation function $\langle G(\vec{r})\rangle$ (center panels) and angle-dependent spatial correlation function $\langle G_{\theta}(\vec{r})\rangle$ (right hand panels) of each image. The average correlation function $\langle G(\vec{r})\rangle$ of all the images reveal local minima and maxima, implying the presence of characteristic length scales for the phase separation. However, this information averages over all directions of space, and the angular dependent correlation function $\langle G_{\theta}(\vec{r})\rangle$ more clearly reveals that there is a particular spatial pattern associated with the phase separation. Inspection of the right hand panels in figure \ref{fig_HRTEM} reveals arcs of intensity with an approximately two-fold rotational symmetry. The arcs are imperfect, but repeat with a fixed periodicity, implying a self-organized pattern of phase separation over remarkably large length scales. Such a pattern of intensity in $\langle G_{\theta}(\vec{r})\rangle$ is consistent with a real space phase separation comprising partially disorderd stripes (see supplemental material). 
In order to quantify the length scales associated with the orthorhombic variation, the average spatial correlation function $\langle G(\vec{r})\rangle$, and the angle-dependent spatial correlation function $\langle G_{\theta}(\vec{r})\rangle$ were computed for each $[$110$]_T$/$[101]_T$ filtered IFFT image (see supplemental material for definitions). %Positions of local mimima in the correlation function of the images studied in this paper provide a measure of the average distance between regions of different crystal structure (orthorhombic or tetragonal), while positions of local maxima provide a measure of the separation between regions of similar crystal structure. In the framework of a phase separated material, we are therefore able to obtain the characteristic length scales of phase separation from these two quantities.
Fig. \ref{fig_HRTEM} shows filtered-and-reconstructed HRTEM images for a representative sample of each Bi composition studied (left panels), after a resolution reduction from 0.47$\AA$ per pixel, to 4.1$\AA$ per pixel, therefore eliminating the atomic resolution information while keeping the longer-range variation in ``orthorhombicity'' (see supplemental material). Both, $\langle G(\vec{r})\rangle$ (shown in supplemental material) and $\langle G_{\theta}(\vec{r})\rangle$ (shown on the right panels of fig. \ref{fig_HRTEM}) of all the images shown, reveal local minima and maxima, implying the presence of characteristic length scales for the phase separation. Furthermore, the angular dependent correlation function $\langle G_{\theta}(\vec{r})\rangle$ clearly reveals that there is a particular spatial pattern associated with the phase separation. Inspection of these quantities, in the right-hand panels of figure \ref{fig_HRTEM}, reveals arcs of intensity with an approximately two-fold rotational symmetry. The arcs are imperfect, but repeat with a fixed periodicity, implying a self-organized pattern of phase separation over remarkably large length scales. Such a pattern of intensity in $\langle G_{\theta}(\vec{r})\rangle$ is consistent with a real space phase separation comprising partially disordered stripes (see supplemental material). 
For a system with stripes separated by a distance $d$ and running along an angle $\alpha$ with respect to the horizontal, the distance between stripes as measured at an angle $\theta$ is given by $N\times d/\cos((\alpha-90^{\circ})-\theta)$ (with $N=1,2,3,...$), which diverges at $\theta=\alpha$. As can be observed in Fig. \ref{fig_HRTEM} (and in similar data shown in the supplemental material), most of the samples studied exhibit this characteristic dependence, with periodic maxima (shown by solid lines in the figure) and minima (dashed lines) that approximately follow such an inverse cosine function. The orientation of the stripes with respect to the crystal axes is not identical for all images studied, but on average it is close to 29$^{\circ}\pm$22$^{\circ}$ from the [100]$_T$ orientation. These stripes are clearly evident in the larger area real space images shown in the left hand panels of Fig. \ref{fig_HRTEM}(a,b), running approximately top-left to bottom-right. In addition to the separation of stripes, inspection of the images in Fig. \ref{fig_HRTEM} reveals that there is a shorter (and more isotropic) length scale of structural variation, which describes the broken-up character of the stripes. This length scale can be seen more clearly in the average correlation function as a kink in the low-r tail, which can be better identified in the derivative of $\langle G(\vec{r})\rangle$. %(see supplemental material). 

\begin{figure}[ht]
\vspace{-0.7cm}
\hspace{-1.2cm}
\centering
\includegraphics[scale=0.36]{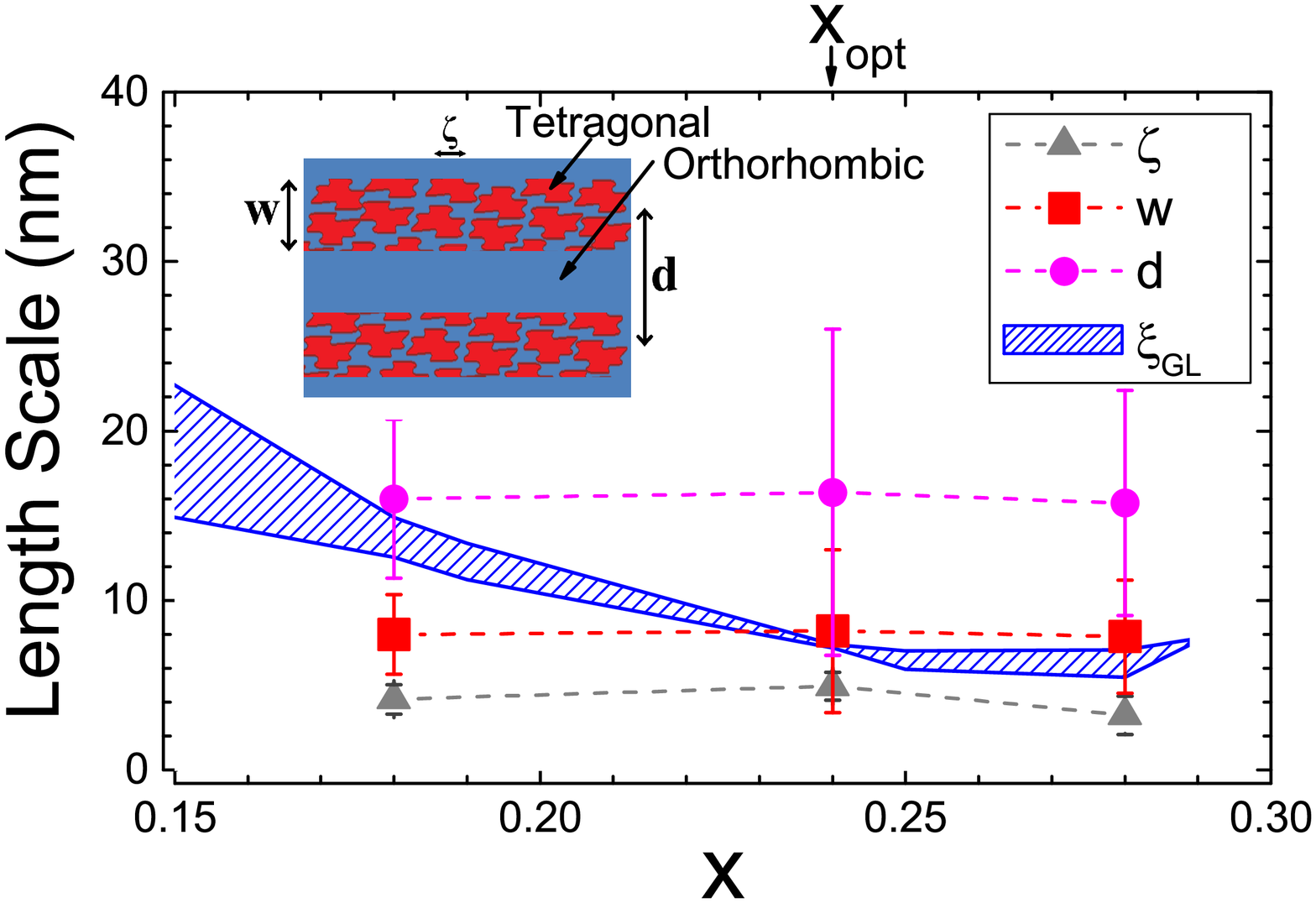}\\% \vspace{-0.3cm}%\hspace{-0.9cm}\\
\vspace{-2cm}
\caption{(Color online) Average characteristic lengths of phase separation for BaPb$_{1-x}$Bi$_x$O$_3$ as a function of $x$. 
%Pink circles represent the average distance between regions of similar orthorhombicity, $d$; red squares represent the width of individual stripes, $w$; and grey triangles represent the correlation length within a single stripe, $\zeta$. Error bars represent the standard error obtained from measurements of different samples. %Lines are drawn between data points to guide the eye. 
These data are contrasted with the Ginzburg-Landau coherence length $\xi_{GL}(0)$, represented in the blue curves.%, calculated from $H_{c2}(T)$ measurements % and determined using a 50$\%$ and 90$\%$ criteria in the resistive superconducting transition (see main text).
}\label{fig_TEMfeatures}
%\vspace{-0.4cm}
\end{figure}

Although the stripe-like character of the structural phase separation is imperfect, nevertheless by identifying the morphology of the nanostructure we are able to define the characteristic length scales of phase separation in terms of three simple parameters (see inset to fig. \ref{fig_TEMfeatures}): the stripe period, $d$, (i.e. the distance between stripes of similar ``orthorhombicity'', determined from the maxima of $\langle G_{\theta}(\vec{r})\rangle$); the stripe width, $w$ (estimated from the regions of minimum values in $\langle G_{\theta}(\vec{r})\rangle$, i.e. stripes half-period, which can be used as a measure of the upper bound to the width of individual stripes); and the length scale associated with disorder within a stripe, $\zeta$, (identified in the derivative of the low-r tail of $\langle G(\vec{r})\rangle$). 
The analysis described above, was performed for a total of five $x=0.18$ samples, four $x=0.24$ samples and four $x=0.28$ samples (all of which are shown in the supplemental material), and the average value of $d$, $w$ and $\zeta$ for each Bi composition were calculated. The results are summarized in Fig. \ref{fig_TEMfeatures}, together with the error obtained by calculating the standard deviation from the average value. 

% ********************************* Cartoon was here.

The phase separation of tetragonal and orthorhombic polymorphs is presumably driven by changes in the relative free energy of the two phases, both as a function of temperature and composition. The resulting morphology is reminiscent of spinodal decomposition%\cite{Ian1}
, but the physical origin is somewhat different in this case, involving two competing phases (see supplemental material). Significantly, in such a scenario, the composition $x_{opt} \sim$ 0.24, at which the tetragonal volume fraction is maximal, marks the separatrix between formation of two different orthorhombic phases, both with the same structure, but one with a lower Bi concentration (for compositions $x < x_{opt}$), and one with a higher Bi concentration ($x > x_{opt}$). Since the free energy of each polymorph can be affected by strain \cite{lee1}, local variations in the local strain is anticipated to broaden or smear the otherwise sharp distinction in the variation of Bi composition. Considering the temperature dependence of the resistivity for compositions that have only an orthorhombic structure, it is clear that Bi substitution leads to a progressive evolution of the electronic properties of the orthorhombic phase from a ``bad metal'' for $x \ll x_{opt}$ (i.e. $d\rho/dT > 0$, but with a very large absolute value of the resistivity) to a ``bad insulator'' for $x \gg x_{opt}$ (i.e. $d\rho/dT < 0$, but nevertheless extrapolating to a finite conductivity at $T$ = 0) \cite{Paula1, Paula2, Uchida1}. It is unclear whether this evolution of the electronic properties of the orthorhombic phase is driven by disorder due to the increasing Bi concentration, or a progressive increase in the CDW correlation length, or indeed a combination of both effects, but tunneling data clearly indicates that the zero temperature conductivity decreases to zero linearly in the entire range from $x =$ 0 to $x =$ 0.3, and that the associated zero bias tunneling anomaly also varies smoothly over this range \cite{Katherine1}. Of particular significance for the following discussion, if the Bi concentration deviates from $x_{opt} \sim$ 0.24 in either direction, and the tetragonal volume fraction correspondingly diminishes, the phase separation results in small islands of superconducting tetragonal material with a characteristic length scale embedded in a matrix of orthorhombic BaPb$_{1-x}$Bi$_x$O$_3$ that is either poorly conducting for $x < x_{opt}$ or poorly insulating for $x > x_{opt}$. This distinction has important consequences for the evolution of the superconducting properties.

The significance of the structural modulation can be readily appreciated by comparing the associated length scales of the disordered stripes, $d$, $w$ and $\zeta$ (solid data points in fig. \ref{fig_TEMfeatures}), with the Ginzburg-Landau coherence length, $\xi_{GL}(0)$ (blue curve in the same figure), for samples with $x < x_{opt}$, $x \approx x_{opt}$ (optimally doped), and $x > x_{opt}$, with approximate $T_c$ values of 7K, 10.5K and 7K respectively. We estimate $\xi_{GL}(0)$ from $H_{c2}(0)$, having used the standard Werthamer-Helfand-Hohenberg approximation to determine $H_{c2}(0)$ from $H_{c2}(T)$. We employed both 50$\%$ and 90$\%$ criteria to extract $H_{c2}(T)$ from resistive transitions, leading to a narrow band of estimated values for $\xi_{GL}(0)$. Inspection of Fig. \ref{fig_TEMfeatures} reveals that the three length scales associated with the phase separation are of the same order of magnitude as the superconducting coherence length, and largely independent of Bi concentration. The shortest length scale, $\zeta$, which characterizes the size of coherent regions within a given stripe, has a weak composition dependence, but does not grow to be larger than the superconducting coherence length for any composition, and is therefore expected to be less relevant than the larger length scales $d$ and $w$ associated with the period and width of the stripes. For low Bi concentrations, $x < x_{opt}$, the coherence length is larger than the width of individual stripes. However, at optimal doping, the width of individual stripes almost exactly matches the superconducting coherence length. Further increasing the Bi concentration appears to result in a saturation of $\xi_{GL}(0)$ which remains comparable to $w$. This behavior is highly suggestive of an important role for the nanostructure in determining the shape of the superconducting dome, as we describe below. 

In the context of an electronically-inhomogeneous system, where the coulomb potential seen by electrons varies spatially in a periodic way, with characteristic length $\lambda$, it has been shown theoretically that $T_c$ does not necessarily track the pairing scale $\Delta_0$, i.e. the superconducting gap magnitude \cite{Kivelson4}. Rather, the evolution of $T_c$ is bounded above by two parameters: the pairing scale $\Delta_0$ and the phase ordering temperature $T_{\theta}$. In the limit where $\lambda\ll\xi$ (where $\xi$ is the superconducting coherence length), $T_{\theta}\gg\Delta_0$, and $T_c$ will be determined by $\Delta_0$. However, in the limit $\lambda\gg\xi$, the phase ordering temperature $T_{\theta}$ is small compared to the pairing amplitude $\Delta_0$, and $T_c$ is entirely determined by $T_{\theta}$, meaning that $T_c$ is suppressed with respect to $\Delta_0$. In this regime the material behaves as a granular superconductor, characterized by superconducting ``islands'' that are only weakly coupled. For a system where the length scale of phase separation evolves with respect to the superconducting coherence length (or vice-versa), the maximum $T_c$ value is obtained in the crossover regime of the curves of $T_{\theta}$ and $\Delta_0$, which happens at $\lambda\sim\xi$. This regime has been dubbed ``optimal inhomogeneity''\cite{Kivelson4,Kivelson7}. In the case of BaPb$_{1-x}$Bi$_x$O$_3$, the phase separation is not necessarily refering to electronic phase separation due to variations in coulomb interaction, but rather the local variation in pairing interaction of the two coexisting polymorphs, although a similar set of arguments clearly applies.

The phenomenology of BaPb$_{1-x}$Bi$_x$O$_3$ appears to be consistent with such a scenario, in which tetragonal and orthorhombic polymorphs correspond to regions of the bulk material with large and small pairing interactions respectively. The evolution with doping of the relative length scales characterized by $\xi_{GL}$ and the phase separation is very suggestive of optimal doping being a turning point from a macroscopic inhomogeneous superconductor (with $\xi_{GL}$ bigger than other characteristic length scales associated with disorder) for $x < x_{opt}$ to a phase-fluctuation-dominated granular superconductor for $x > x_{opt}$ (illustrated schematically in figure \ref{fig_BPBOmodel}). Indeed, several signatures of granular superconductivity are observed in this regime, such as negative magnetoresistance for fields above $H_{c2}(T)$, and scaling reminiscent of a superconductor-insulator quantum phase transition \cite{Paula1}. Additionally, scanning tunneling spectroscopy (STS) measurements for compositions beyond optimal doping show a large variation in gap values as a function of position, with maximum values exceeding those found in the higher $T_c$ optimally doped material \cite{hari1}, suggesting that samples with $x > x_{opt}$ have a larger local pairing amplitude than expected for their macroscopic $T_c$, and even for an 11 K superconductor. This observation is consistent with a macroscopic $T_c$ being bounded by the phase ordering line, $T_{\theta}$, i.e., with a granular superconductor picture (see figure \ref{fig_BPBOmodel}). Significantly, these observations imply that for $x > x_{opt}$ the superconducting phase of this material (the tetragonal polymorph) is in fact a higher-temperature superconductor, possibly even comparable to the other bismuthate superconductor Ba$_{1-x}$K$_x$BiO$_3$ \cite{Katherine1}.

In the above analysis, the only significance of the stripe-like character of the  nanostructure of BaPb$_{1-x}$Bi$_x$O$_3$ has been that it has enabled us to establish the characteristic length-scales with a little more precision than if we had assumed a more isotropic morphology. However, the stripe-like morphology possibly has a much deeper significance. In the context of BaPb$_{1-x}$Bi$_x$O$_3$, this might provide a natural means to understand the unusual scaling behavior observed at the superconductor-insulator transition close to optimal doping in this material \cite{Paula1,Paula2}, motivating theoretical investigation of percolation effects near the quantum phase transition for a material with a ``stripy'' morphology. More broadly, several families of underdoped cuprates have been shown to exhibit stripe and/or unidirectional CDW formation \cite{Letacon, Tranquada1, Howald2}. In the case of the cuprates, the stripe/CDW order is driven by spontaneous electronic order, whereas for the bismuthates the stripe-like nanostructure is quenched from higher temperature \cite{Cavap}. Significantly, in both cases, the stripe-like phase separation and superconductivity are found to have comparable length scales. In this broader context, BaPb$_{1-x}$Bi$_x$O$_3$ provides a model system to explore the effects of stripe-like phase separation on superconductivity, and in particular on the associated phenomenology of optimal inhomogeneity.

The authors thank S. A. Kivelson for helpful discussions. This work is supported by AFOSR Grant No. FA9550-09-1-0583. The electron microscopy was performed at Ames Laboratory (Y.Z. and M.J.K.) and supported by the U.S. Department of Energy (DOE), Office of Basic Energy Science (BES), Division of Materials Sciences and Engineering, under Contract No. DE-AC02-07CH11358. C.P. and H.C.M. were supported by US Department of Energy (DOE), Office of Science, Basic Energy Sciences (BES), Materials Sciences and Engineering Division, under contract DE-AC02-76SF00515.

%%%%%%%%%%%%%%%%%%%%%%%%%%%%%%%%%%%%%%%%%%%%%%%%%%%%%%%%%%
%%%%%%%%%%%%%%%%%%%%%% APPENDIX %%%%%%%%%%%%%%%%%%%%%%%%%%
\let\clearpage\relax
\appendix
\section*{Appendices}
\setcounter{figure}{0} 
\makeatletter 
\renewcommand{\thefigure}{A\@arabic\c@figure}
\makeatother
%\tableofcontents

%\renewcommand{\thesection}{A-\arabic{equation}}

%\setcounter{figure}{0} 
%\makeatletter 
%\renewcommand{\thefigure}{A\@arabic\c@figure}
%\makeatother

\section{Glazer's notation}

The ideal cubic perovskite ABO$_3$, described by the space group $Pm\bar{3}m$, can be represented as a network of corner-sharing BO$_6$ octahedra. `A' atoms sit in the geometric center of the gap between oxygen octahedra. This structure is a ``simple'' and highly symmetric one; however, most materials with perovskite structures are not in their ideal cubic form, but their structure can nevertheless be represented as coming from distortions from this ideal configuration. The types of distortions found in perovskites can be narrowed down to three types: B-cation displacements within an octahedra; distortions of the BO$_6$ octahedral unit; and, the most common one and subject of this section, and of Glazer's study \cite{Glazer}, the rigid tilting of the corner-sharing BO$_6$ linked-octahedra units. This last type of distortion was described by Glazer in terms of tilt components along the three different pseudocubic (PC) axes, referred to the original undistorted cubic perovskite. Such pseudocubic axes coincide with the tetrad axes of the octahedra. Given the octahedra corner connections, a tilt about a pseudocubic axis determines the tilts in the directions perpendicular to this axis. However, the tilt of the successive octahedra along the same axis can be either in the same direction or in the opposite direction. With this in mind, the different possibilities of tilt-distortions can be labeled by the notation $a^*b^*c^*$, where $a$, $b$, $c$ refer to tilts around the $[100]_{PC}$, $[010]_{PC}$ and $[001]_{PC}$ axes, respectively. If letters are repeated, the tilts are equal for their respective axis. The superscript $*$ can be either $0$, for no-tilt along an axis; $+$, for tilt of successive octahedra in the same sense; or $-$, for tilt of successive octahedra in the opposite sense \cite{Howard}. For example, the \textit{I4/mcm} space group is represented by the notation $a^0a^0c^-$, which means zero tilt about the $[100]_{PC}$  and $[010]_{PC}$ axes, and finite tilt about the $[001]_{PC}$ axis, with opposite rotation of the successive octahedra along this axis. The \textit{Ibmm} space group is represented by the notation $a^-a^-c^0$, which means equal tilts about the $[100]_{PC}$ and $[010]_{PC}$ axes (equivalent to a tilt about the $[110]_{PC}$ direction), with opposite rotation of the successive octahedra along these axes, and zero tilt about the $[001]_{PC}$ axis.

\section{Electron diffraction patterns}

Simulated electron diffraction patterns for tetragonal \textit{I4/mcm} and orthorhombic \textit{Ibmm} polymorphs of BaPb$_{1-x}$Bi$_x$O$_3$ along the $[001]_T$ and $[010]_T$ zone axis were obtained through the University of Illinois web-based electron microscopy application software (WEB-EMAPS) \cite{emaps}, using the atomic parameters shown in tables \ref{table_tetra} and \ref{table_ortho}. Along both of these zone axis, the $(hkl)$ set of reflections with $hlk$ even, are common to both, orthorhombic and tetragonal phases. However, for both zone axes, the $(hkl)$ set of reflections with $hlk$ odd, appears only in the orthorhombic phase and not in the tetragonal. In ref. \citenum{Paula1} we showed that it is possible to recreate the spatial separation of the two polymorphs by systematically masking these diffraction peaks and performing an inverse Fourier transform (IFFT). The result of applying a mask such that only the even $(hkl)$ peaks, common to both the tetragonal and orthorhombic phases, show an ordered array of planes of atoms (see figure 6 in ref. \citenum{Paula1}). In contrast, the result of applying a mask to the $[101]_T/[110]_T$ peaks, attributed only to the \textit{Ibmm} orthorhombic phase, reveals a spatial variation due to the densely intergrown nanostructure, as shown in Fig. 3 of the main manuscript, and Figs. \ref{fig_Gr2825}, \ref{fig_Gr2790} and \ref{fig_Gr3124} of this supplemental material.

%With this it is possible to recreate the spatial separation of the two polymorphs by systematically masking these diffraction peaks and performing an inverse Fourier transform (IFFT). 

%The odd $\{hkl\}$ series of reflections appear don't appear in the tetragonal phase (i.e., $\{110\}$ or $\{101\}$, for $[001]$ and $[010]$ zone axis, respectively), and are exclusive of the orthorhombic polymorph, whereas the even $\{hkl\}$ series of reflections (i.e., $\{200\}$ and $\{220\}$, or $\{200\}$ and $/\{202\}$, for $[001]$ and $[010]$ zone axis, respectively) appear in both tetragonal and orthorhombic phases. In this way, we can it is possible to recreate the spatial separation of the two polymorphs by systematically masking these diffraction peaks and performing an inverse Fourier transform (IFFT)

\begin{table}[!htbp]
\begin{tabular}{ | l | c | c | c |  c | c |  }
%\begin{tabular}{ | l  c  c   c  c   c  c |  }
  \hline                       
  Atom & Wyck. & Site & x/a & y/b & z/c \\
	\hline \hline
  Ba & 4b & -42m & 0 & 0.5 & 0.25 \\ 
  Pb/Bi & 4c & 4/m & 0 & 0 & 0 \\ 
	O1 & 8h & m.2m & 0.2179(14) & 0.7179(14) & 0 \\  
	O2 & 4a & 422 & 0 & 0 & 0.25 \\ 
  \hline  
\end{tabular}
\caption{Atomic parameters for tetragonal BaPb$_{1-x}$Bi$_x$O$_3$ (space group \textit{I4/mcm}, No. 140), with $x\approx 0.28$, as reported in ref. \citenum{Cava1}. The site 4c is fully ocupied by Pb/Bi, and the Pb to Bi ratio is determined by $x$.}\label{table_tetra}
\end{table}

\begin{table}[!htbp]
\begin{tabular}{ | l | c | c | c |  c | c |  }
%\begin{tabular}{ | l  c  c   c  c   c  c |  }
  \hline                       
  Atom & Wyck. & Site & x/a & y/b & z/c \\
	\hline \hline
  Ba & 4e & mm2 & 0.496 & 0 & 0.25 \\ 
  Pb/Bi & 4a & 2/m & 0 & 0 & 0 \\ 
	O1 & 4e & mm2 & 0.0496 & 0 & 0.25 \\  
	O2 & 8g & .2. & 0.25 & 0.25 & 0.9741 \\ 
  \hline  
\end{tabular}
\caption{Atomic parameters for orthorhombic BaPb$_{1-x}$Bi$_x$O$_3$ (space group \textit{Ibmm}, No. 74), with $x\approx 0.28$, as reported in ref. \citenum{Cava1}. The site 4a is fully ocupied by Pb/Bi, and the Pb to Bi ratio is determined by $x$.}\label{table_ortho}
\end{table}

\section{Definition of the correlation function}

The spatial autocorrelation function $G(\vec{r})$ of an image is defined as the statistical correlation of two points separated by a vector $\vec{r}=\vec{r}_i-\vec{r}_j$, where $\vec{r}_i$ and $\vec{r}_j$ are the positions of those two points in the image \cite{bianconi}.

\begin{equation}
G(\vec{r})=\frac{1}{N(\vec{r})}\sum_{i,j}\frac{(I_i-\langle I\rangle_1)(I_j-\langle I\rangle_2)}{\sigma_1\sigma_2}
\end{equation}

where

\begin{subequations}
\begin{align}
N(\vec{r})&=\sum_{i,j}\delta_{\vec{r}, (\vec{R}_i-\vec{R}_j)}\\
\langle I\rangle_{1}&=\frac{1}{N(\vec{r})}\sum_{i,j}\delta_{\vec{r}, (\vec{R}_i-\vec{R}_j)} I_{i}\\
\langle I\rangle_{2}&=\frac{1}{N(\vec{r})}\sum_{i,j}\delta_{\vec{r}, (\vec{R}_i-\vec{R}_j)} I_{j}\\
\sigma_{1}^2 &=\left(\frac{1}{N(\vec{r})}\sum_{i,j}\delta_{\vec{r}, (\vec{R}_i-\vec{R}_j)} I_{i}^2\right)-\left(\langle I\rangle_{1}\right)^2\\
\sigma_{2}^2 &=\left(\frac{1}{N(\vec{r})}\sum_{i,j}\delta_{\vec{r}, (\vec{R}_i-\vec{R}_j)} I_{j}^2\right)-\left(\langle I\rangle_{2}\right)^2
\end{align}
\end{subequations}

The average spatial autocorrelation function $\langle G(\vec{r})\rangle$ is the result of averaging the correlation function of all vectors with the same magnitude $|\vec{r}|$. The angle-dependent autocorrelation function $\langle G_{\theta}(\vec{r})\rangle$ is the result of averaging the correlation function of all vectors with orientation $\theta$ with respect to the horizontal axis, and magnitude $|\vec{r}|$.

\begin{figure}[!htbp]
%\vspace{-1cm}
\centering
\hspace{-0.7cm}
\includegraphics[scale=0.31]{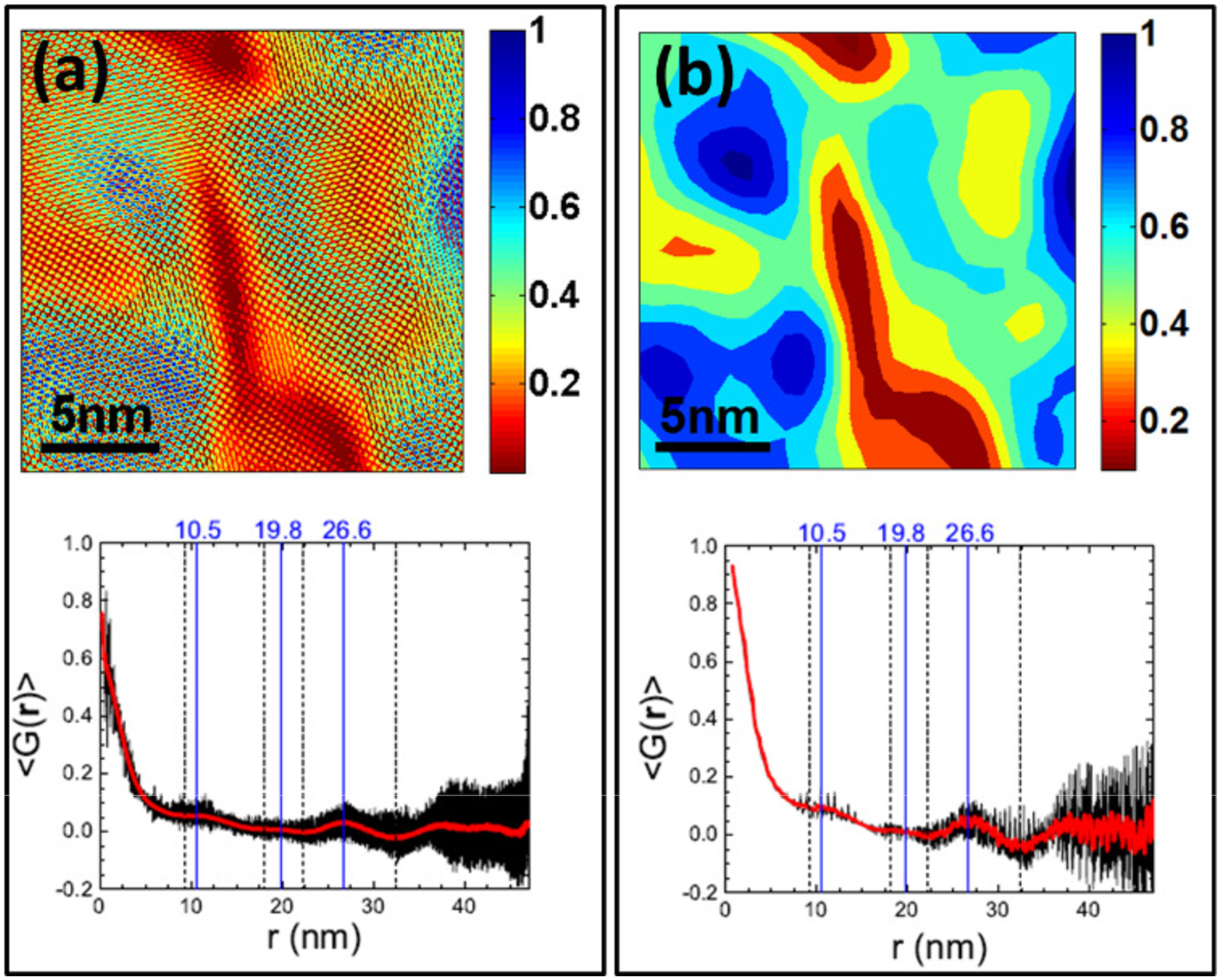} \hspace{-0.6cm}
%\includegraphics[scale=0.4]{2825_2_440Kx_abs_300to700.eps} \hspace{-0.6cm}
%\includegraphics[scale=0.4]{2825_2_440Kx_abs_redresolby16_19to44.eps} \\ \vspace{-0.4cm}\hspace{-0.7cm}
%\includegraphics[scale=0.47]{correlation_2825-2-440Kx_redby4_original_andSmooth.eps} \hspace{-0.5cm}
%\includegraphics[scale=0.47]{correlation_2825-2-440Kx_redby16_original_andSmooth.eps} \\
%\vspace{-0.3cm}
\caption{(Color online) {\bf (a)} 19$\times$19nm$^2$ portion of a $[110]_T$ filtered-and-reconstructed HRTEM image for a sample with bismuth concentration of $x$=0.18, and its corresponding correlation function below it, showing the atomic resolution detail. {\bf (b)} Same image as in (a), and its corresponding correlation function, after a 7.5$\AA\times$7.5$\AA$ averaging, eliminating the atomic resolution information while maintaining the broader orthorhombic structural variation. The red curves in the correlation functions are 3nm smoothing of the original black curves.}\label{fig_redresol}
%\vspace{-0.4cm}
\end{figure}

\begin{figure}[!htbp]
%\vspace{-0.5cm}
\hspace{1cm}
\centering
\hspace{-1.03cm}
\hspace{1cm}
\includegraphics[scale=0.73]{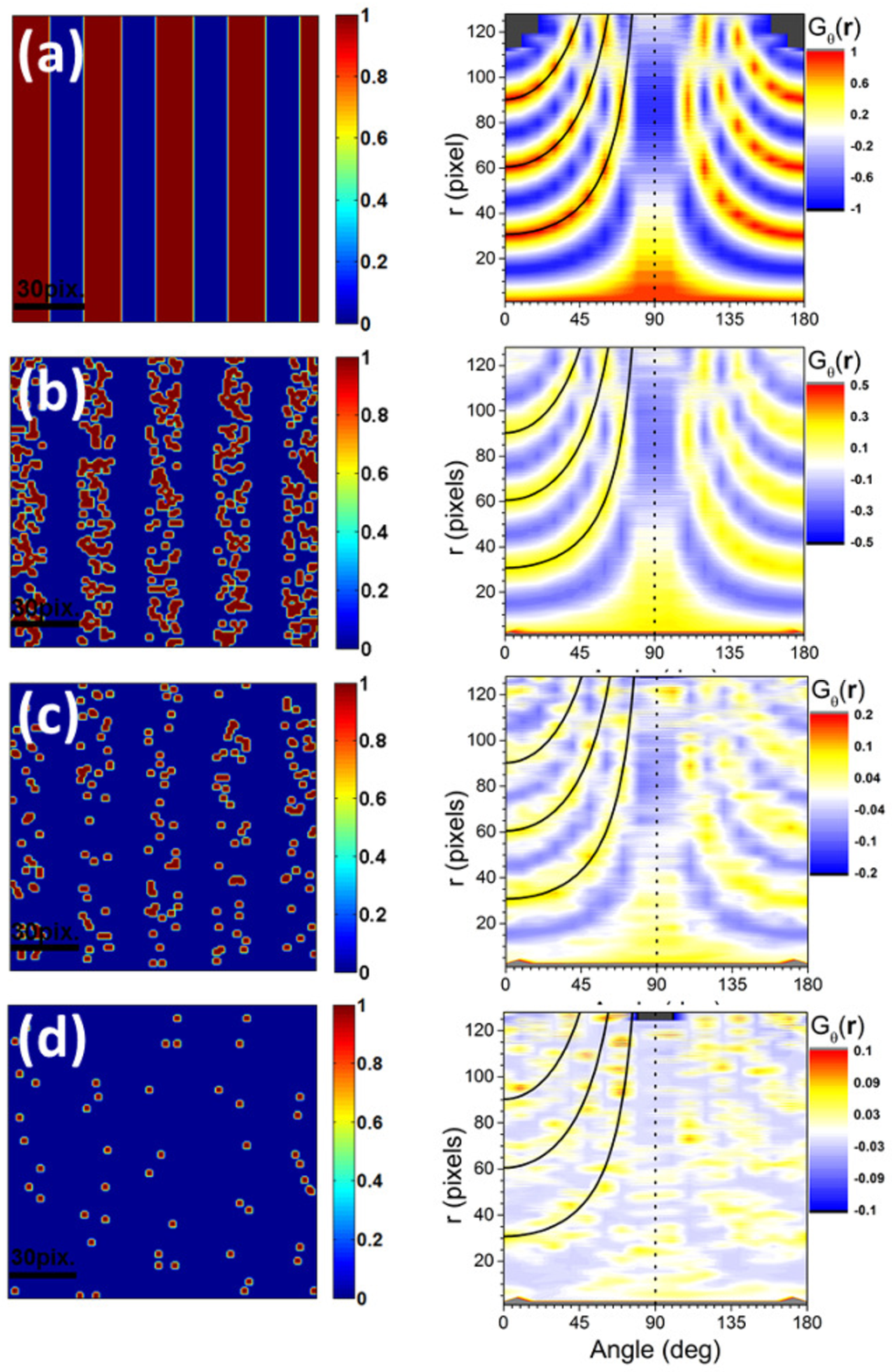} \\ %\vspace{-0.1cm}\hspace{-0.6cm}
%\vspace{-0.3cm}
\caption{(Color online) Images of 128$\times$128 pixels, with stripes of width $w=$15 pixels, separated between them by $d=$30 pixels. For the different images, a broken-up character of a different level was introduced, as a number of islands of size 3$\times$3 pixels, placed at random positions within the red stripes. The number of islands within a stripe for each image is: \textbf{(a)} Infinite (perfect stripe formation), \textbf{(b)} 120 islands, \textbf{(c)} 40 islands, \textbf{(d)} 10 islands. For each image, the angle-dependent correlation function $\langle G_{\theta}(\vec{r})\rangle$ is shown in the right-hand panel. Black lines in these plots follow the functional form $N*d/\cos((\alpha-90^{\circ})-\theta)$, where $N=1,2,3,...$, $d=$30 and $\alpha=$90.}\label{fig_stripes}
%\vspace{-0.4cm}
\end{figure}

\begin{figure*}[!htbp]
%\vspace{-0.5cm}
\hspace{1cm}
\centering
\hspace{-2cm}
\hspace{1cm}
\includegraphics[scale=1.1]{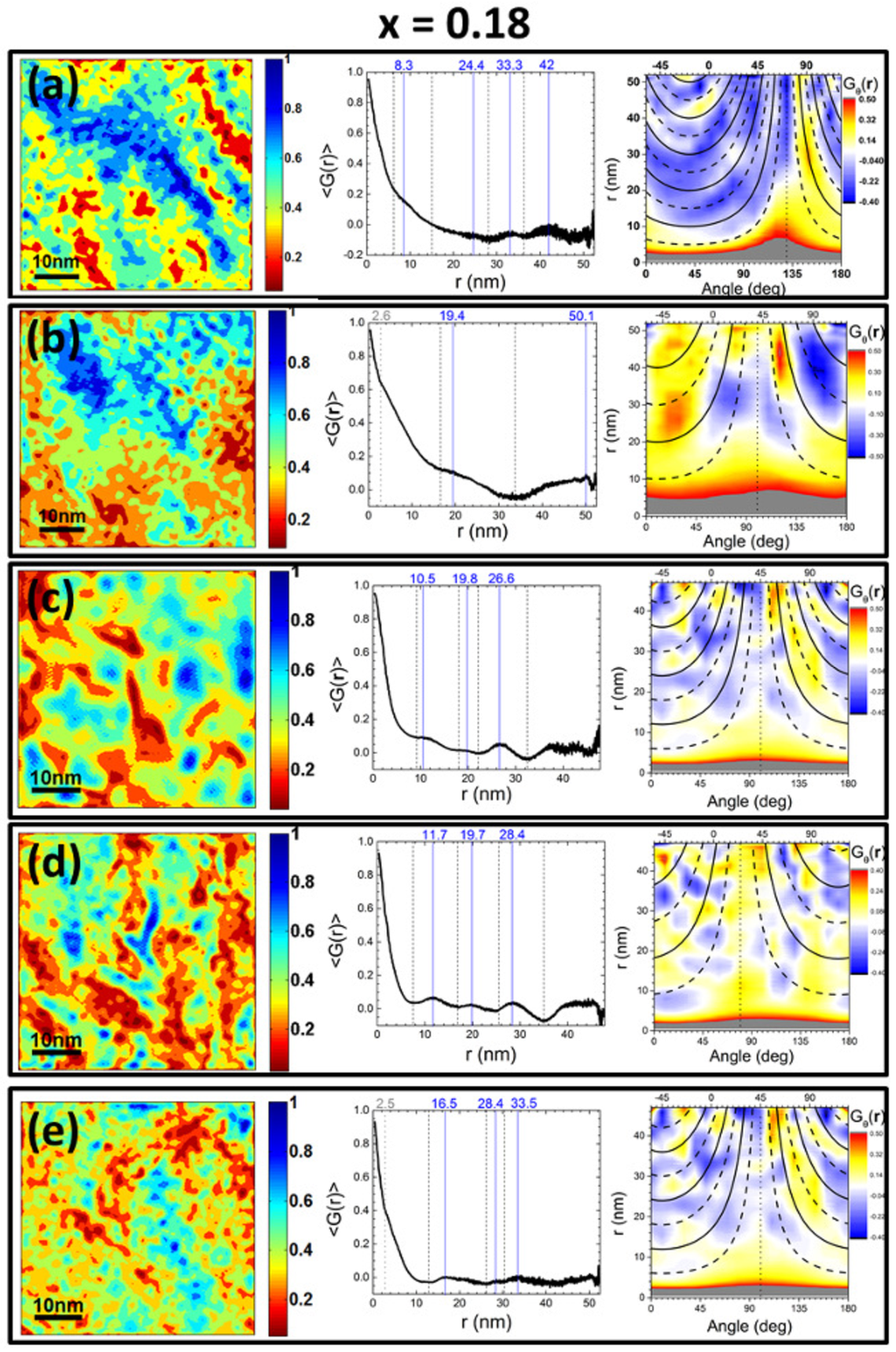} \\ %\vspace{-0.1cm}\hspace{-0.6cm}
%\vspace{-0.3cm}
\caption{(Color online) $[110]_T$/$[101]_T$ filtered-and-reconstructed HRTEM images (first-column figures), for different samples with bismuth concentration of $x$=0.18. For each horizontal panel, the center column shows the corresponding average spatial correlation function, $\langle G(\vec{r})\rangle$. Solid blue vertical lines indicate the local maxima in $\langle G(\vec{r})\rangle$, while dashed-black vertical lines indicate local minima. The third column in each horizontal panel shows the angle-dependent spatial correlation function, $\langle G_{\theta}(\vec{r})\rangle$ on a color scale, as a function of $|\vec{r}|$ (vertical axis) and the angle $\theta$ with the horizontal (bottom-axis) or the $\left[200\right]_T$ crystalline axis (top-axis). Solid and dashed lines represent the best fits to $N\times d/\cos((\alpha-90^{\circ})-\theta)$ and $(2N-1)\times w/\cos((\alpha-90^{\circ})-\theta)$ for the local maxima and minima respectively.}\label{fig_Gr2825}
%\vspace{-0.4cm}
\end{figure*}

\begin{figure*}[!htbp]
\vspace{0.5cm}
\hspace{-1cm}
\centering
\hspace{-1cm}
\hspace{1cm}
\includegraphics[scale=1.1]{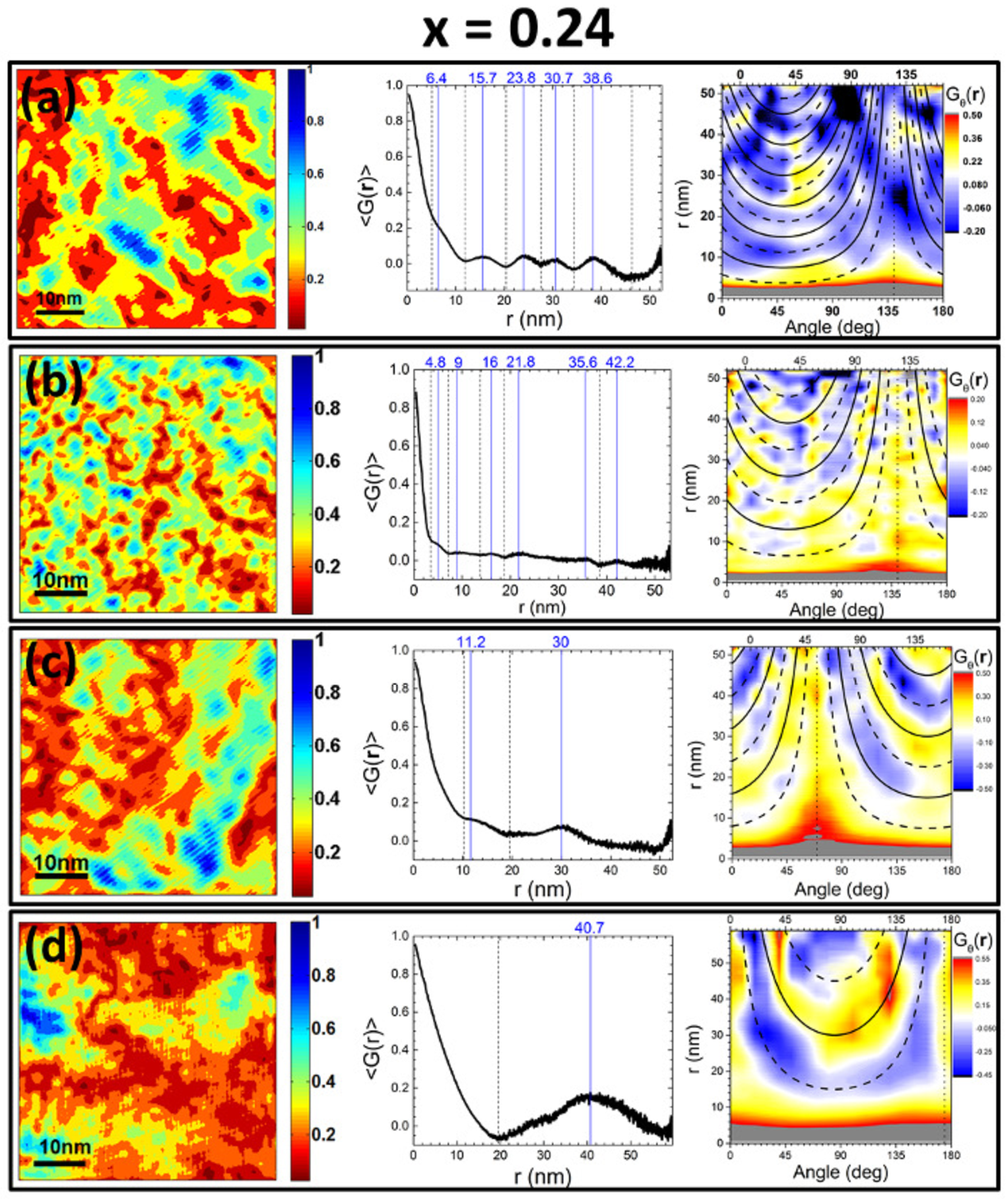} \\ %\vspace{-0.1cm}\hspace{-0.6cm}
\vspace{0.5cm}
\caption{(Color online) $[110]_T$/$[101]_T$ filtered-and-reconstructed HRTEM images (first-column figures), for different samples with bismuth concentration of $x$=0.24. For each horizontal panel, the center column shows the corresponding average spatial correlation function, $\langle G(\vec{r})\rangle$. Solid blue vertical lines indicate the local maxima in $\langle G(\vec{r})\rangle$, while dashed-black vertical lines indicate local minima. The third column in each horizontal panel shows the angle-dependent spatial correlation function, $\langle G_{\theta}(\vec{r})\rangle$ on a color scale, as a function of $|\vec{r}|$ (vertical axis) and the angle $\theta$ with the horizontal (bottom-axis) or the $\left[200\right]_T$ crystalline axis (top-axis). Solid and dashed lines represent the best fits to $N\times d/\cos((\alpha-90^{\circ})-\theta)$ and $(2N-1)\times w/\cos((\alpha-90^{\circ})-\theta)$ for the local maxima and minima respectively.}\label{fig_Gr2790}
\vspace{0.5cm}
\end{figure*}

\begin{figure*}[!htbp]
%\vspace{-0.5cm}
%\hspace{1cm}
\centering
\hspace{-1.03cm}
\hspace{1cm}
\includegraphics[scale=1.2]{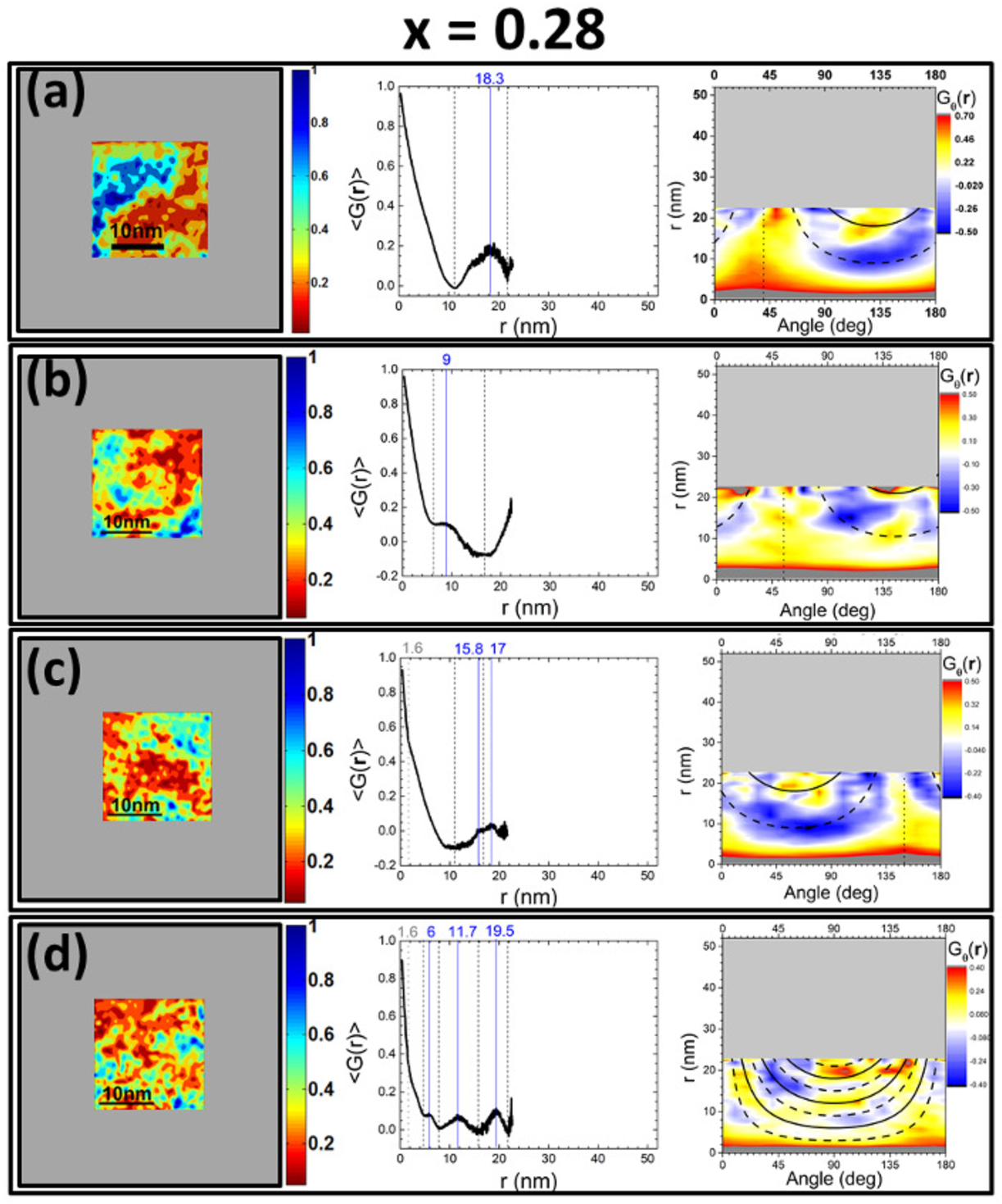} \\ %\vspace{-0.1cm}\hspace{-0.6cm}
%\vspace{-0.3cm}
\caption{(Color online) $[110]_T$/$[101]_T$ filtered-and-reconstructed HRTEM images (first-column figures), for different samples with bismuth concentration of $x$=0.28. These images were taken for a smaller area than the ones for the $x=$0.18 and $x=$0.24 samples. To preserve a direct comparison with those data, the images are shown on the same scale as for figs. \ref{fig_Gr2825} and \ref{fig_Gr2790}. For each horizontal panel, the center column shows the corresponding average spatial correlation function, $\langle G(\vec{r})\rangle$. Solid blue vertical lines indicate the local maxima in $\langle G(\vec{r})\rangle$, while dashed-black vertical lines indicate local minima. The third column in each horizontal panel shows the angle-dependent spatial correlation function, $\langle G_{\theta}(\vec{r})\rangle$ on a color scale, as a function of $|\vec{r}|$ (vertical axis) and the angle $\theta$ with the horizontal (bottom-axis) or the $\left[200\right]_T$ crystalline axis (top-axis). Solid and dashed lines represent the best fits to $N\times d/\cos((\alpha-90^{\circ})-\theta)$ and $(2N-1)\times w/\cos((\alpha-90^{\circ})-\theta)$ for the local maxima and minima respectively.}\label{fig_Gr3124}
%\vspace{-0.4cm}
\end{figure*}

\begin{figure*}[!ht]
\vspace{-0.5cm}
\hspace{-0.6cm}
\centering
\includegraphics[scale=0.24]{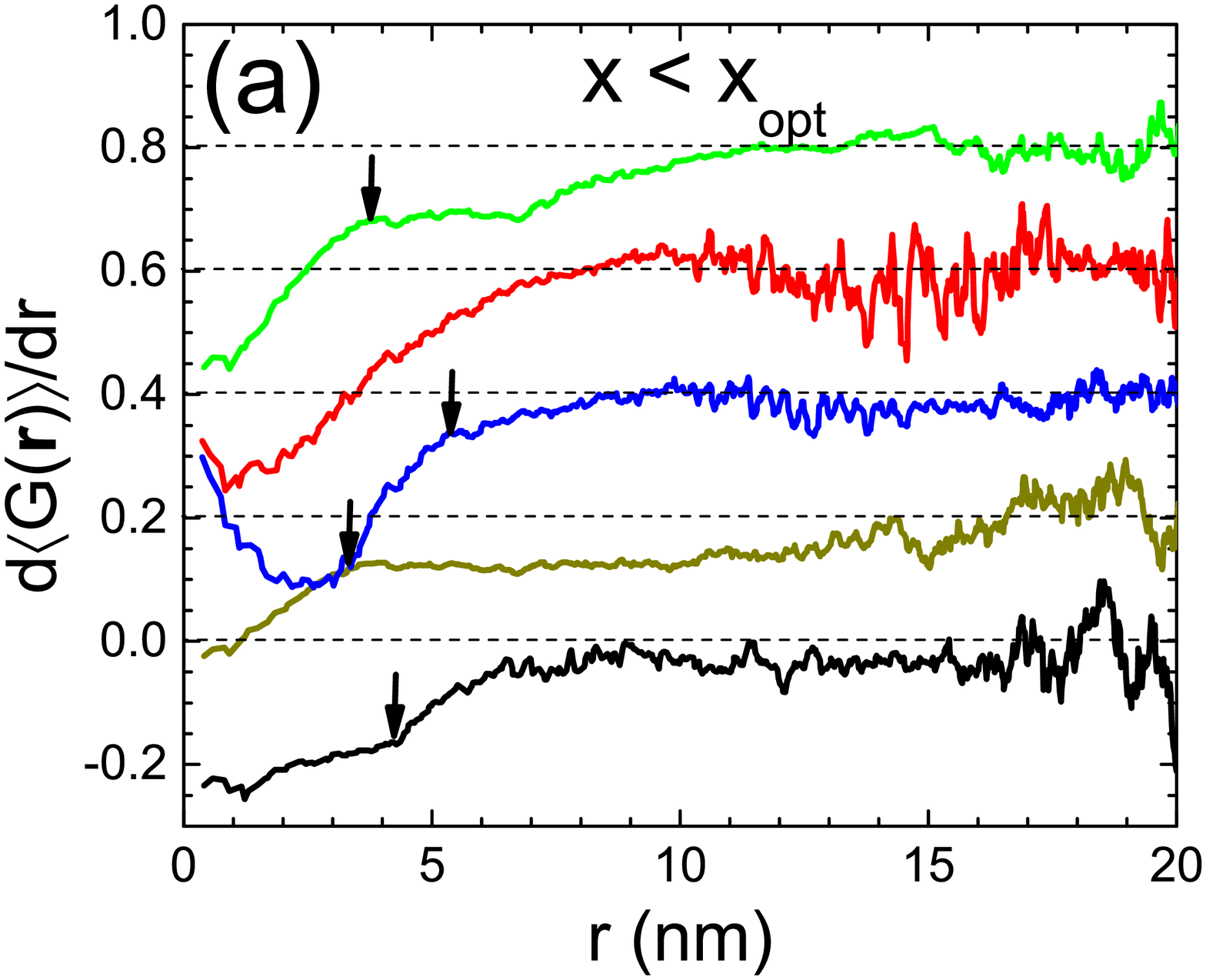} \hspace{-0.9cm}
\includegraphics[scale=0.24]{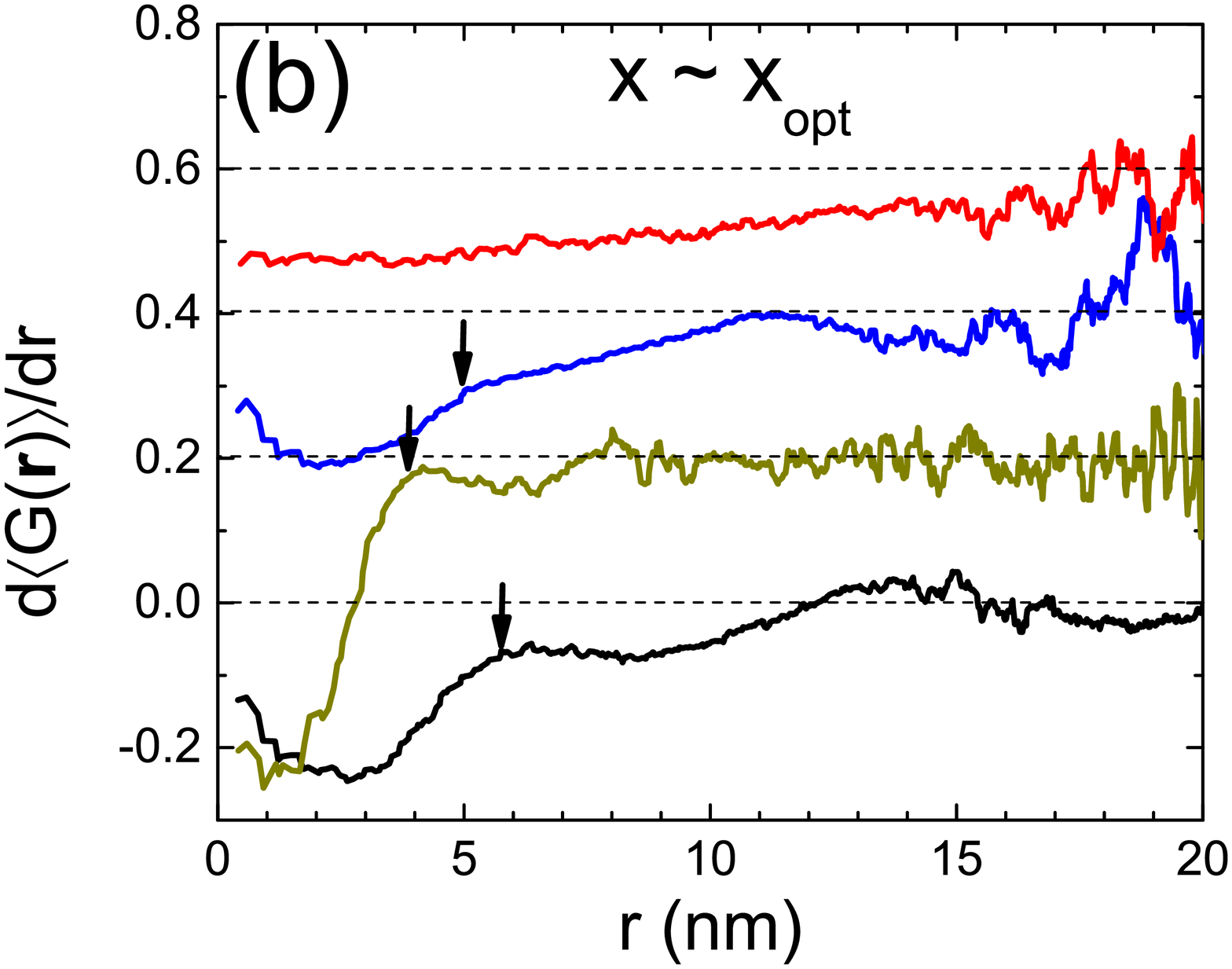}\hspace{-0.9cm}
\includegraphics[scale=0.24]{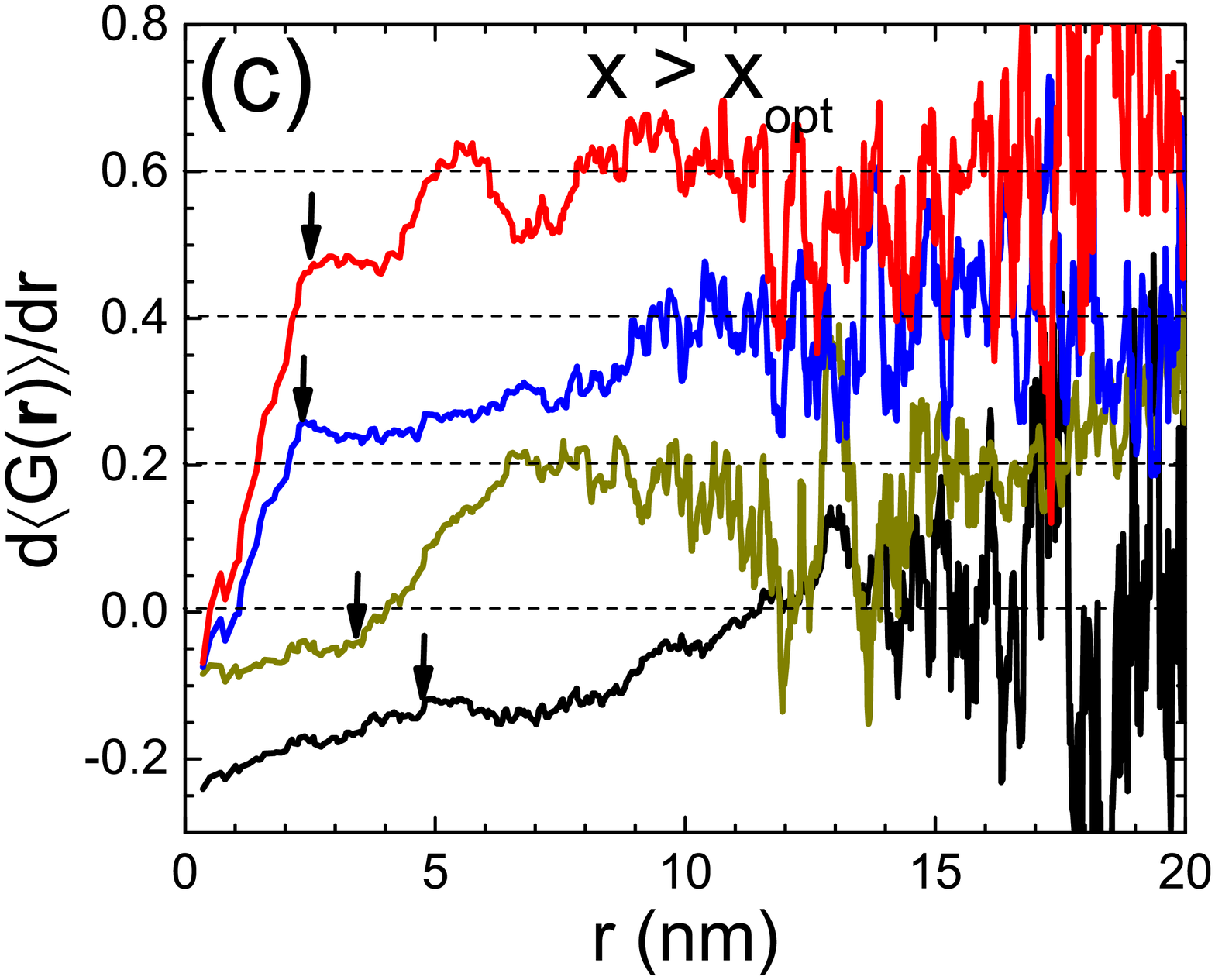} %\hspace{-0.7cm} 
%\vspace{-0.3cm}
\caption{(Color online) Derivative of the average correlation function $\langle G(\vec{r})\rangle$ (for the low-r tail region) of different filtered-and-reconstructed images of samples with Bi concentrations of {\bf (a)} $x=$0.18 ($x < x_{opt}$), {\bf (b)} $x=$0.24 ($x \approx x_{opt}$) and {\bf (c)} $x=$0.28 ($x > x_{opt}$). Dashed horizontal lines mark the $d\langle G(\vec{r})\rangle/dr=0$ zone for each curve. Arrows mark the points determining the disorder length-scale within a stripe, $\zeta$. Average values of $\zeta$ are shown in Fig. 4 of the main manuscript, as a function of $x$. }\label{fig_derivative}
%\vspace{-0.4cm}
\end{figure*}

\section{Correlation function for reduced-resolution images}

%Fig. \ref{fig_redresol}(a) shows a 19x19nm$^2$ portion of the $[101]_T$ filtered IFFT of the HRTEM image in fig. 1(b) of the main manuscript (with bismuth concentration of $x$=0.18), showing the atomic resolution detail. Fig. \ref{fig_redresol}(b) shows the same image in (a) after a 7.5$\AA$x7.5$\AA$ averaging, therefore eliminating the atomic resolution information while maintaining the broader orthorhombic structural variation. The bottom parts of fig. \ref{fig_redresol}(a) and \ref{fig_redresol}(b) show the computed average spatial correlation functions of their respective images on top. The vertical lines in $\langle G(\vec{r})\rangle$ label local minima and maxima positions $|\vec{r}|$, being equivalent for both, the original resolution image in fig. \ref{fig_redresol}(a), and the reduced resolution image in fig. \ref{fig_redresol}(b). For the analysis presented in the main manuscript we considered only the reduced resolution images, given that these conserve the information of the longer-scale structural variation while reducing the computational requirements for the spatial correlation function calculation.

Fig. \ref{fig_redresol}(a) shows a 19$\times$19nm$^2$ portion of the $[$110$]_T$ filtered IFFT of the HRTEM image in fig. 2(b) of the main manuscript. This image keeps the information of both the atomic periodicity as well as a larger-scale contrast variation, reflecting variations in the local ``orthorhombicity'' across the sample. The image in figure \ref{fig_redresol}(b) is the result of a resolution reduction by adjacent averaging, of the image in fig. \ref{fig_redresol}(a), from 0.47$\AA$ per pixel, to 7.5$\AA$ per pixel, therefore eliminating the atomic resolution information while keeping the longer-range variation in ``orthorhombicity''. The bottom parts of fig. \ref{fig_redresol}(a) and \ref{fig_redresol}(b) show the computed average spatial correlation functions of their respective images on top. The vertical lines in $\langle G(\vec{r})\rangle$ label local minima and maxima positions $|\vec{r}|$, being equivalent for both, the original resolution image in fig. \ref{fig_redresol}(a), and the reduced resolution image in fig. \ref{fig_redresol}(b). For the purpose of our analysis, we consider only the reduced resolution images, given that these conserve the information of the longer-scale structural variation while reducing the computational requirements.

\section{Correlation function for a stripe model}

The angular dependent correlation function $\langle G_{\theta}(\vec{r})\rangle$ was computed for an image of size 128$\times$128 pixels, showing perfect stripes formation, with stripes of width $w=$15 pixels and periodicity $d=$30 pixels, running along an angle $\alpha=$90$^{\circ}$ with respect to the horizontal axis (fig. \ref{fig_stripes}(a)). The color scale of the right hand side plot of fig. \ref{fig_stripes}(a) represents the value of $\langle G_{\theta}(\vec{r})\rangle$, as a function of the angle $\theta$ with the horizontal axis and the magnitude of $\vec{r}$. Maxima of $\langle G_{\theta}(\vec{r})\rangle$ for this image appear along arcs following the functional form $N*d/\cos((\alpha-90^{\circ})-\theta)$ (shown by the black solid lines), where $N=1,2,3,...$. Figures \ref{fig_stripes}(b)-(d) show images of the same size and with stripes of the same width and periodicity as in (a), but where a progressively broken-up character has been introduced for each image. For these images, the maxima of $\langle G_{\theta}(\vec{r})\rangle$ follow in average the same functional form as the original zero-disorder stripe model, but the local maximum value of $\langle G_{\theta}(\vec{r})\rangle$ progressively decreases in value, from 1 for the perfect-stripes image in fig. \ref{fig_stripes}(a), to about 0.1 for the most broken-up image in fig. \ref{fig_stripes}(d). At the same time, the arcs where the maximum values of $\langle G_{\theta}(\vec{r})\rangle$ appear get progressively more broken-up, although its average functional form is preserved, and its periodicity can be well identified. The angle dependent correlation functions $\langle G_{\theta}(\vec{r})\rangle$ observed in the $[110]_T/[101]_T$ filtered-and-reconstructed images analyzed throughout this article show very similar features to the ones observed in this model of broken-up stripes.

\begin{figure}[!ht]
%\vspace{-0.5cm}
\hspace{1cm}
\centering
\hspace{-1.03cm}
\hspace{1cm}
\includegraphics[scale=0.35]{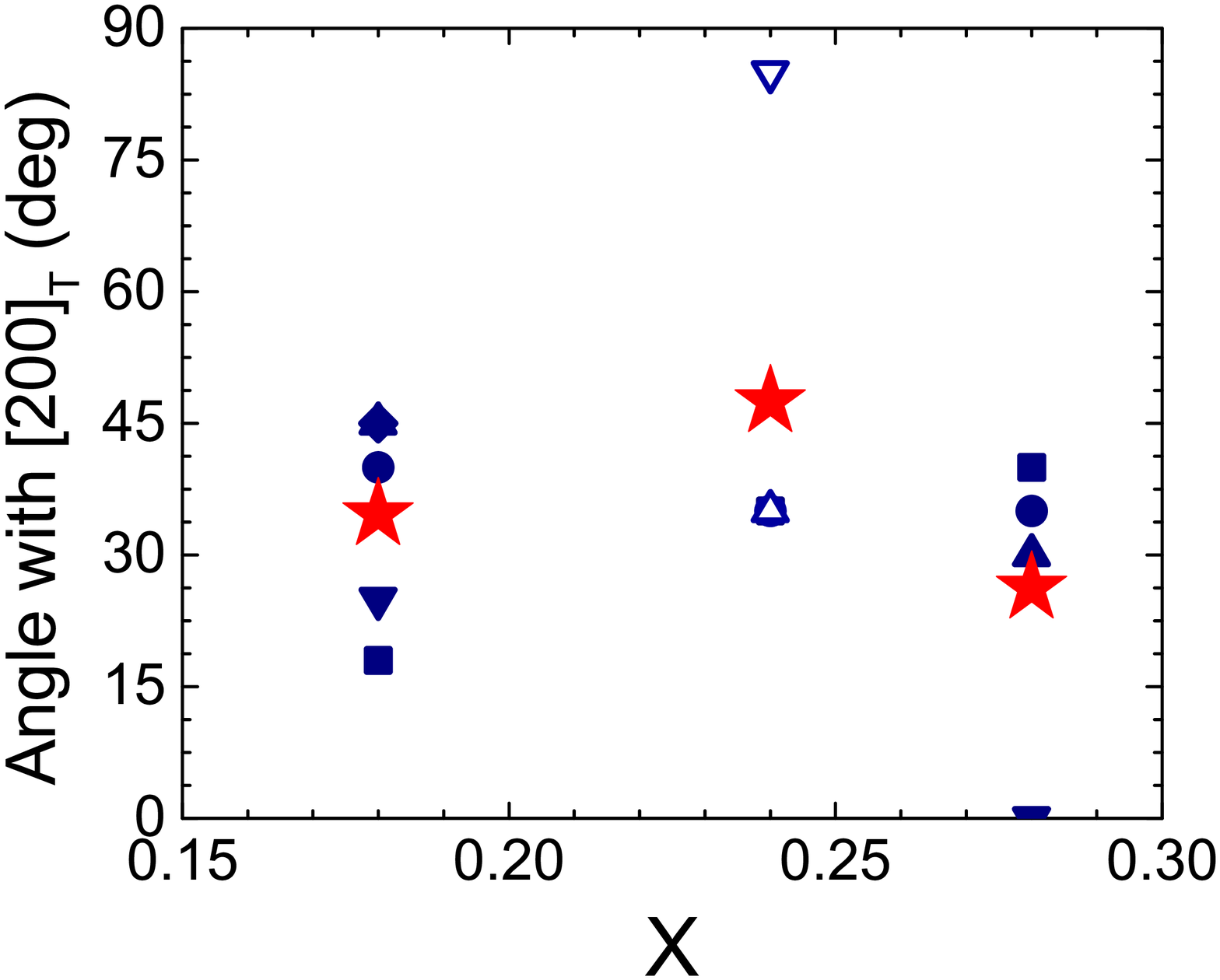} \\ %\vspace{-0.1cm}\hspace{-0.6cm}
\vspace{-0.3cm}
\caption{(Color online) Orientation of stripes with respect to the $\{200\}_T$ crystalline axis as a function of Bi concentration, for all the samples studied. Open blue symbols show the orientation for samples with images taken along the $[001]_T$ zone axis; full blue symbols show the orientation for samples with images taken along the $[010]_T$ zone axis. Star-shape full red symbols show the average orientation for each composition. }\label{fig_orientation}
%\vspace{-0.4cm}
\end{figure}

\section{Correlation function for all the images studied}

%The average spatial correlation function $\langle G(\vec{r})\rangle$ and angle-dependent spatial correlation function $\langle G_{\theta}(\vec{r})\rangle$ were computed for extra four $x=0.18$ samples (fig. \ref{fig_Gr2825} in this supplemental material), three $x=0.24$ samples (fig. \ref{fig_Gr2790}) and three $x=0.28$ samples (fig. \ref{fig_Gr3124}), in addition to the ones of each that we showed in figure 2 of the main manuscript. The characteristic length scales of phase separation: stripes periodicity $d$, stripes width $w$ and correlation length within a single stripe, $\zeta$, were determined for these figures (plus the ones shown in the main text), and the average and dispertion for each quantity were determined. Results are shown in Fig. 4 of the main manuscript.

Figures \ref{fig_Gr2825}, \ref{fig_Gr2790} and \ref{fig_Gr3124} of this supplemental material show $[110]_T/[101]_T$ filtered-and-reconstructed images (left-hand panels) for a total of five $x=0.18$ samples (fig. \ref{fig_Gr2825}), four $x=0.24$ samples (fig. \ref{fig_Gr2790}) and four $x=0.28$ samples (fig. \ref{fig_Gr3124}), as well as their respective average spatial correlation function $\langle G(\vec{r})\rangle$ (central panels) and angle-dependent spatial correlation function $\langle G_{\theta}(\vec{r})\rangle$ (right-hand panels). The first image of each figure had been already shown in figure 3 of the main manuscript; however it is shown again for completeness and with the spirit of presenting the average correlation function that was not presented before. The characteristic length scales of phase separation shown in figure 4 of the main manuscript are: (1) stripes periodicity, $d$, (2) stripes width, $w$, and (3) the correlation length within a single stripe, $\zeta$. These length scales were determined for all of the figures, and the average value for each quantity determined and plotted as a function of Bi concentration in Fig. 4 of the main manuscript.

%The average spatial correlation function $\langle G(\vec{r})\rangle$ and angle-dependent spatial correlation function $\langle G_{\theta}(\vec{r})\rangle$ were computed for a total of five $x=0.18$ samples (fig. \ref{fig_Gr2825} in this supplemental material), four $x=0.24$ samples (fig. \ref{fig_Gr2790}) and four $x=0.28$ samples (fig. \ref{fig_Gr3124}). The first image in each figure corresponds to the one already shown in figure 2 of the main manuscript. For all the images, the The characteristic length scales of phase separation: stripes periodicity $d$, stripes width $w$ and correlation length within a single stripe, $\zeta$, were determined for these figures, and the average and dispertion for each quantity were determined. Results are shown in Fig. 4 of the main manuscript.

\begin{figure}[!ht]
%\vspace{-0.5cm}
\hspace{1cm}
\centering
\hspace{-1.03cm}
\hspace{1cm}
\includegraphics[scale=0.8]{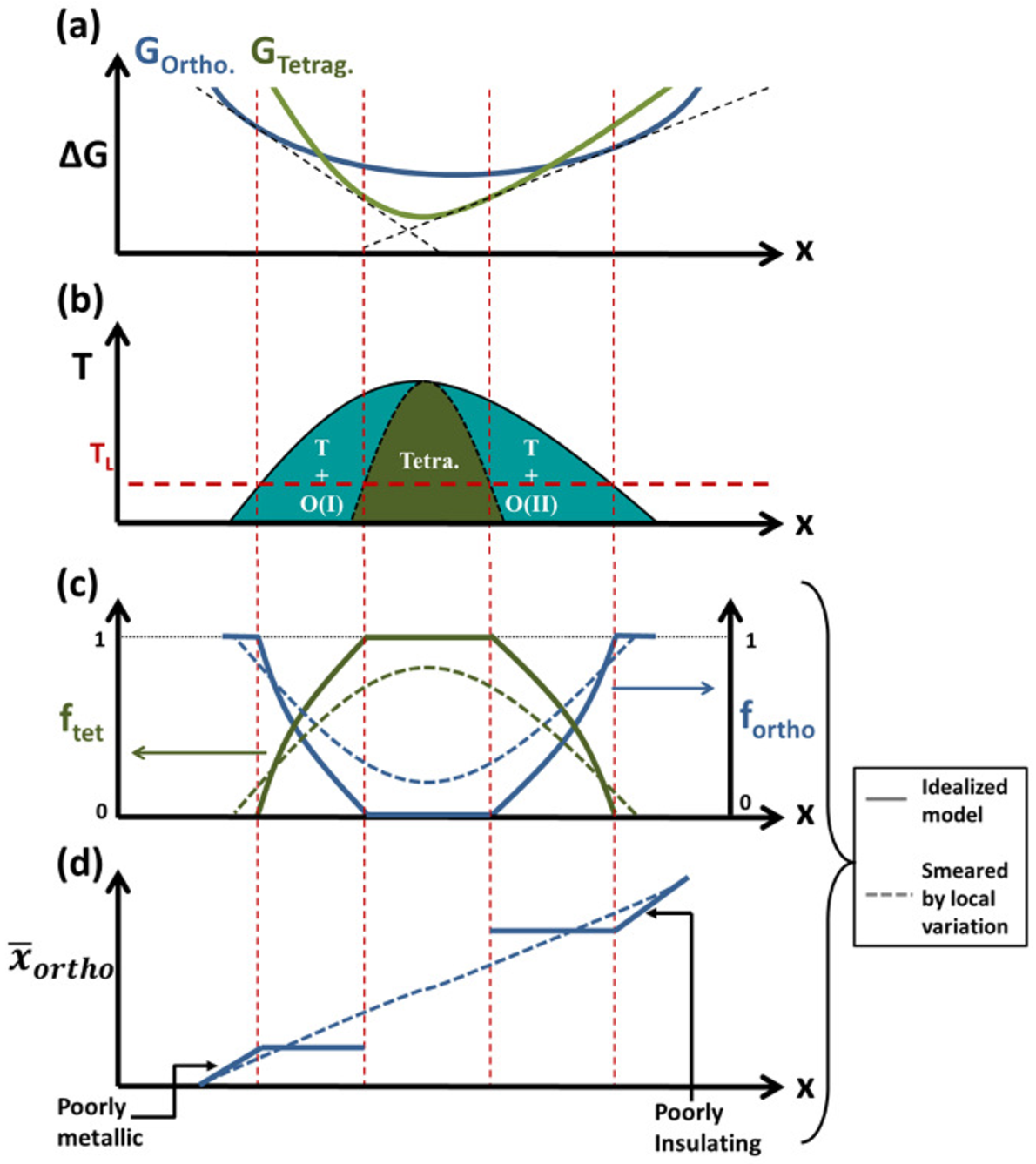} \\ %\vspace{-0.1cm}\hspace{-0.6cm}
\caption{(Color online) \textbf{(a)} Schematic diagram of the Gibbs free energy of orthorhombic (G$_{ortho.}$) and tetragonal (G$_{tetrag.}$) phases as a function of composition at a given low temperature $T=T_L$. \textbf{(b)} The corresponding phase diagram, showing regions of phase coexistence. At $x=x_{opt}$ the tetragonal fraction is maximum. For $x<x_{opt}$ the orthorhombic phase is labeled as O(I), and it is a low-Bi phase, presumably metallic. For $x>x_{opt}$ the orthorhombic phase is labeled O(II), and it is a rich-Bi phase, presumably insulating. \textbf{(c)} Fraction of tetragonal ($f_{tet}$, left-axis) and orthorombic ($f_{ortho}$, right-axis) phases as a function of composition, at temperature $T=T_L$ (marked by the red horizontal line in panel (b)). \textbf{(d)} Spatial average of orthorhombic composition, $\bar{x}_{ortho}$, as a function of composition, for the same temperature $T=T_L$. Solid lines in panels (c) and (d) show the evolution of these quantities with composition for the ideal case depicted in (a), whereas dashed-curves show the evolution for the case where the local free energy is modified by local strain, giving rise to a distribution of free energies that smear the otherwise sharp distinctions and features of the tetragonal and orthorhombic fractions, and spatial average of orthorhombic composition. }\label{fig_spinodal}
%\vspace{-0.3cm}
%\vspace{-0.4cm}
\end{figure}

\begin{figure}[!ht]
%\vspace{-0.7cm}
\hspace{-0.2cm}
\centering
\includegraphics[scale=0.32]{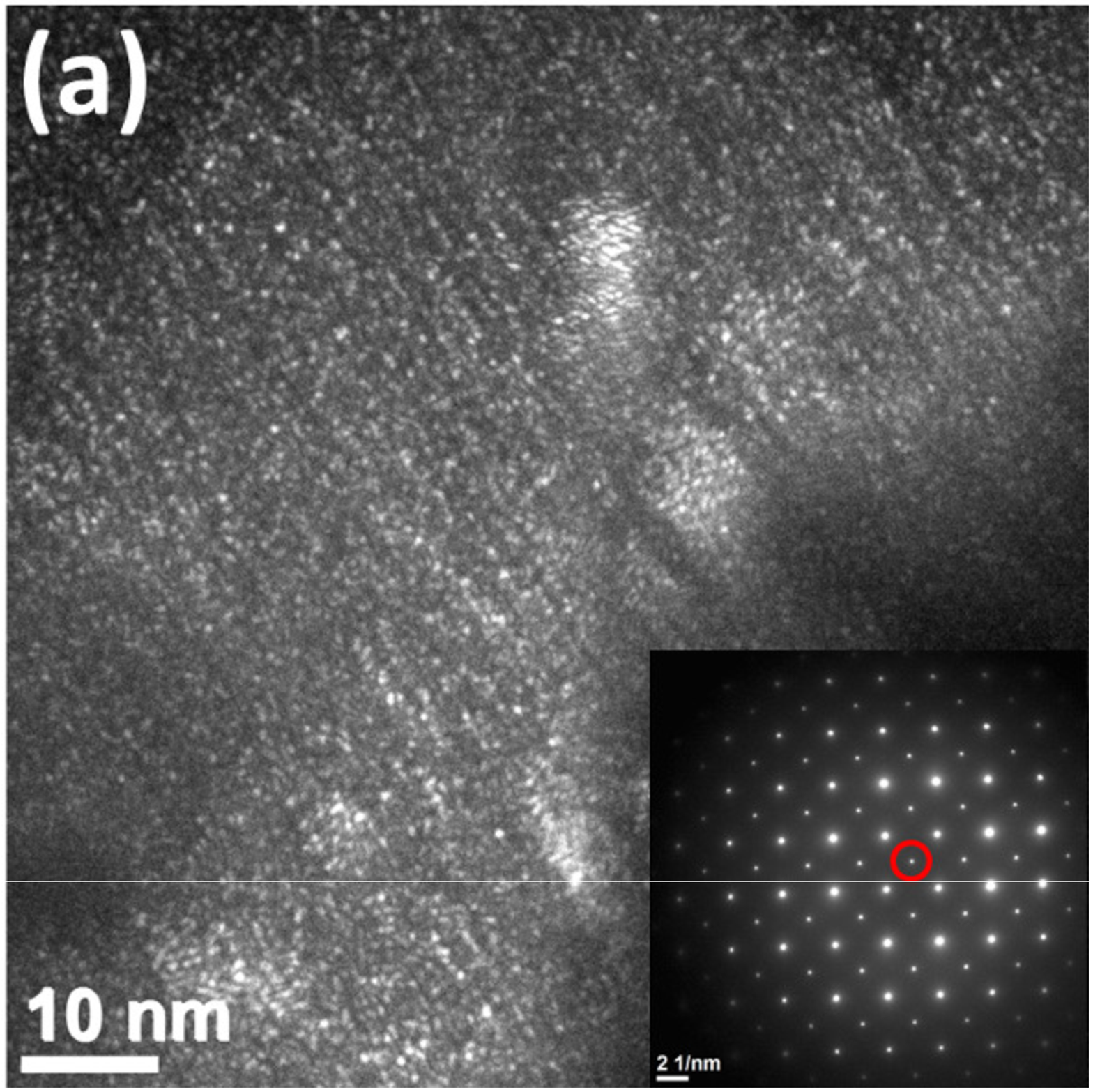} \hspace{-0.5cm} 
\includegraphics[scale=0.315]{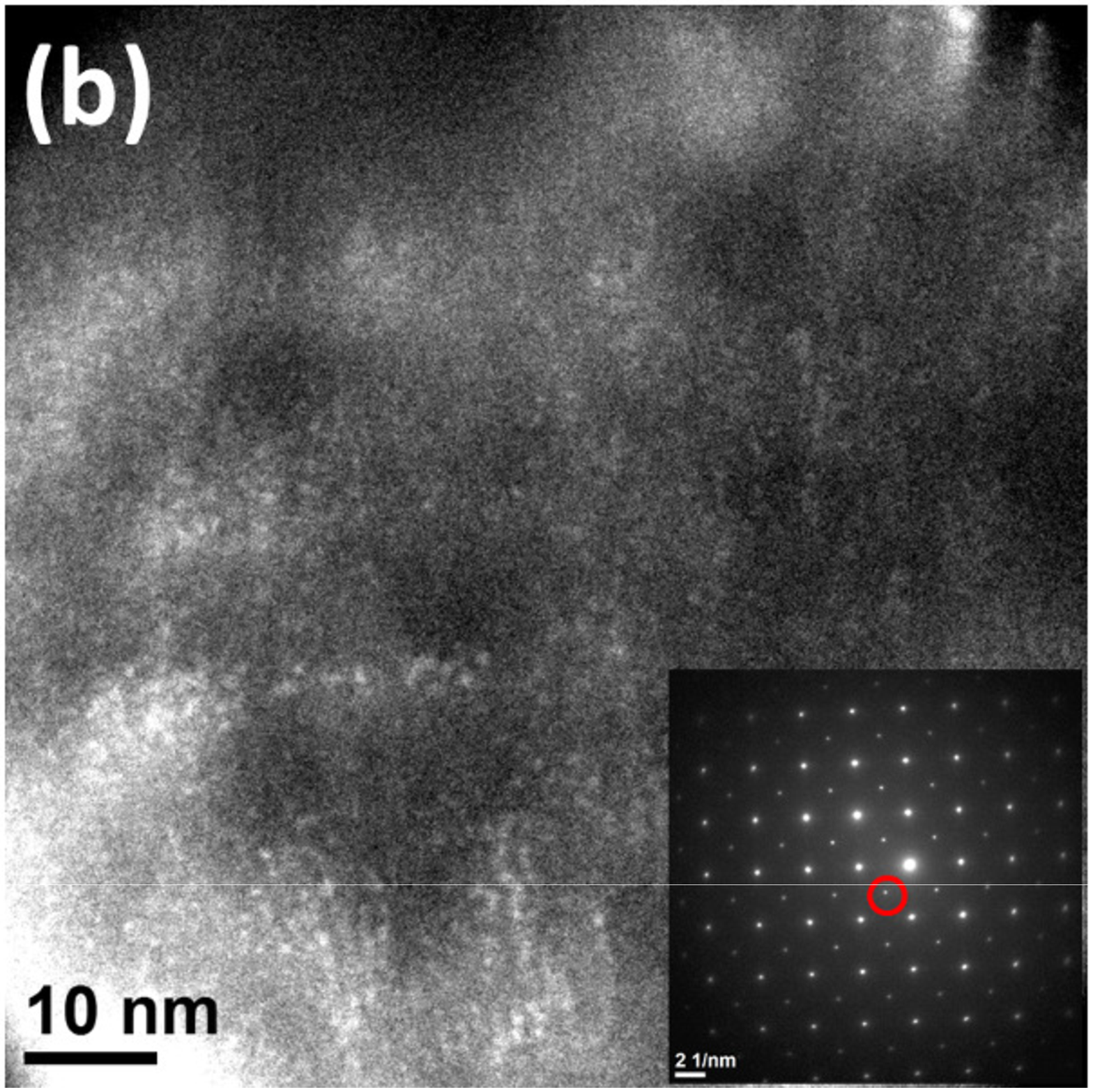}
%\vspace{-0.3cm}
\caption{(Color online) 80.8$\times$80.8 nm$^2$ dark-field TEM images, as looking down the $[001]_T$ zone axis, obtained using the $(110)$ reflection (shown in red in the insets to both figures) for samples with Bi concentration of {\bf (a)} $x=$0.18 and {\bf (b)} $x=$0.28. }\label{fig_dark}
%\vspace{-0.4cm}
\end{figure}

\section{Derivative of the average correlation functions}

The length-scale representing the disorder within a stripe, denoted as $\zeta$, is picked-up more clearly in the average correlation function $\langle G(\vec{r})\rangle$, as a kink or change of slope in the low-r tail region. This kink is more precisely seen in the derivative of this quantity, as shown in figure \ref{fig_derivative}(a,b,c) for the different Bi compositions and samples of each compositions studied. Black arrows in these plots show the points determining the value of $\zeta$ for each sample. The average value of $\zeta$ is plotted in figure 4 of the main manuscript, together with the other length scales of phase separation. %Although the derivative plots show that there could be other features representing the disorder length scale within a stripe, these appear at length scales below the average width of a stripe (around 8nm) for each Bi concentration.

\section{Orientation of stripes}

Fig. \ref{fig_orientation} shows the orientation of stripes with respect to the $[200]_T$ axis, for all the different samples studied, as a function of Bi concentration. The uncertainty of this measure is large given the imperfect character of the stripe patterns, however, it can be observed that the average value is close to 30$^{\circ}$ from the $\{200\}_T$ axis (29$^{\circ}\pm$22$^{\circ}$).

\begin{figure*}[!htbp]
%\vspace{-0.5cm}
\hspace{1cm}
\centering
\hspace{-1.03cm}
\hspace{1cm}
\includegraphics[scale=0.6]{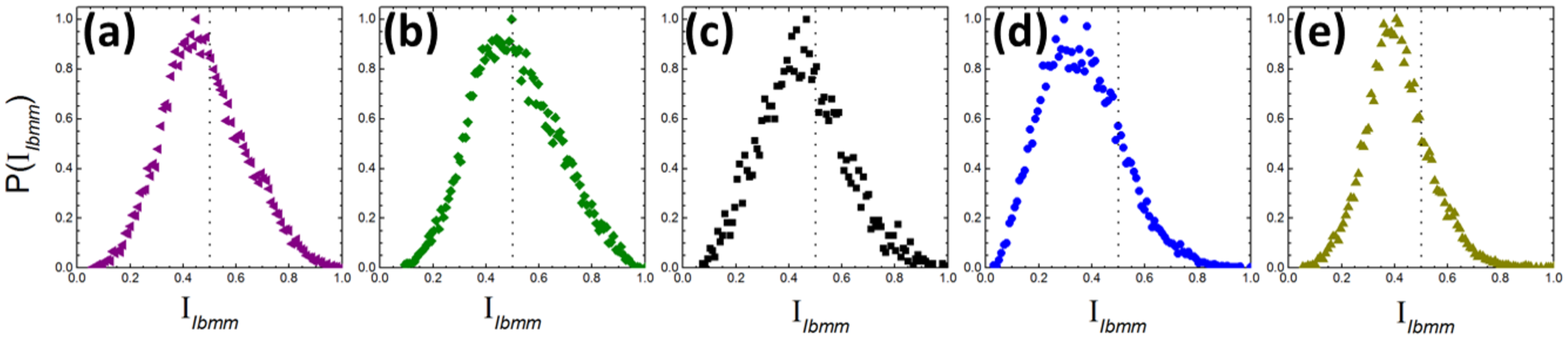} \\ %\vspace{-0.1cm}\hspace{-0.6cm}
\vspace{-0.3cm}
\caption{(Color online) Probability distribution of orthorhombicity, $P(I_{Ibmm})$, as a function of the orthorhombicity intensity, $I_{Ibmm}$, for their corresponding images in figure \ref{fig_Gr2825} (with Bi composition of $x=$0.18). %(e) corresponds to the image in figure 3(a) of the main manuscript. Bi concentration for these images is $x$=0.18. The vertical dashed lines label the point of half intensity in the curves.
}\label{fig_Portho2825}
%\vspace{-0.4cm}
\end{figure*}

\begin{figure*}[!htbp]
%\vspace{-0.5cm}
\hspace{1cm}
\centering
\hspace{-1.03cm}
\hspace{1cm}
\includegraphics[scale=0.6]{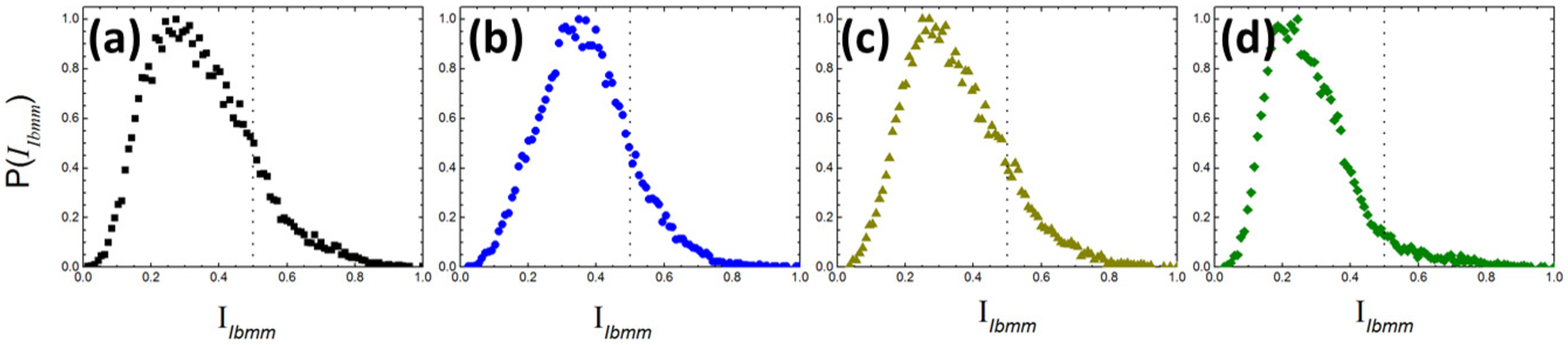} \\ %\vspace{-0.1cm}\hspace{-0.6cm}
\vspace{-0.3cm}
\caption{(Color online) Probability distribution of orthorhombicity, $P(I_{Ibmm})$, as a function of the orthorhombicity intensity, $I_{Ibmm}$, for their corresponding images in figure \ref{fig_Gr2790} (with Bi composition of $x=$0.24). %(d) corresponds to the image in figure 3(b) of the main manuscript. Bi concentration for these images is $x$=0.24. The vertical dashed lines label the point of half intensity in the curves. 
}\label{fig_Portho2790}
%\vspace{-0.4cm}
\end{figure*}

\begin{figure*}[!htbp]
%\vspace{-0.1cm}
\hspace{1cm}
\centering
\hspace{-1.03cm}
\hspace{1cm}
\includegraphics[scale=0.6]{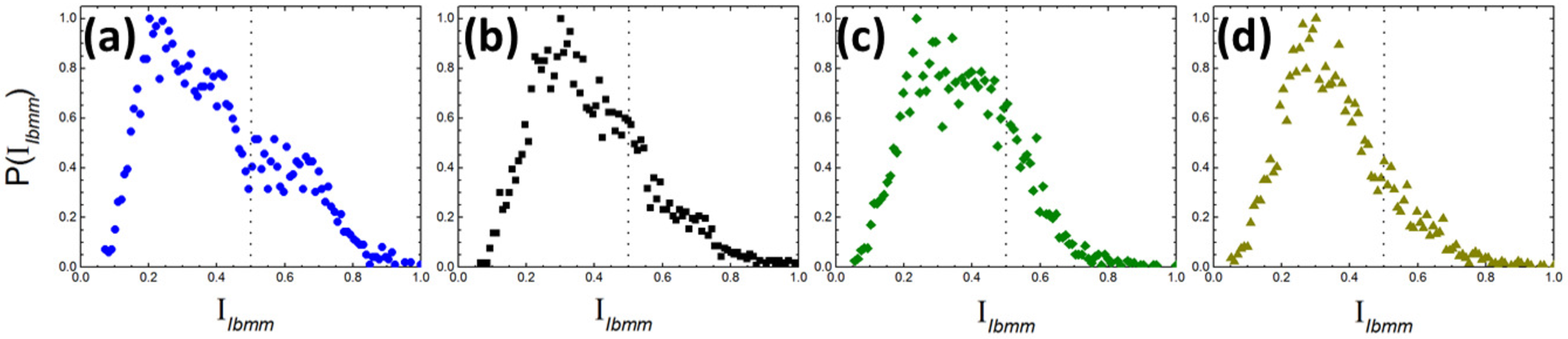} \\ %\vspace{-0.1cm}\hspace{-0.6cm}
\vspace{-0.3cm}
\caption{(Color online) Probability distribution of orthorhombicity, $P(I_{Ibmm})$, as a function of the orthorhombicity intensity, $I_{Ibmm}$, for their corresponding images in figure \ref{fig_Gr3124} (with Bi composition of $x=$0.28). %(d) corresponds to the image in figure 3(c) of the main manuscript. Bi concentration for these images is $x$=0.28. The vertical dashed lines label the point of half intensity in the curves. 
}\label{fig_Portho3124}
%\vspace{-0.4cm}
\end{figure*}

\begin{figure*}[!htbp]
%\vspace{-0.5cm}
\hspace{-0.6cm}
\centering
\includegraphics[scale=0.24]{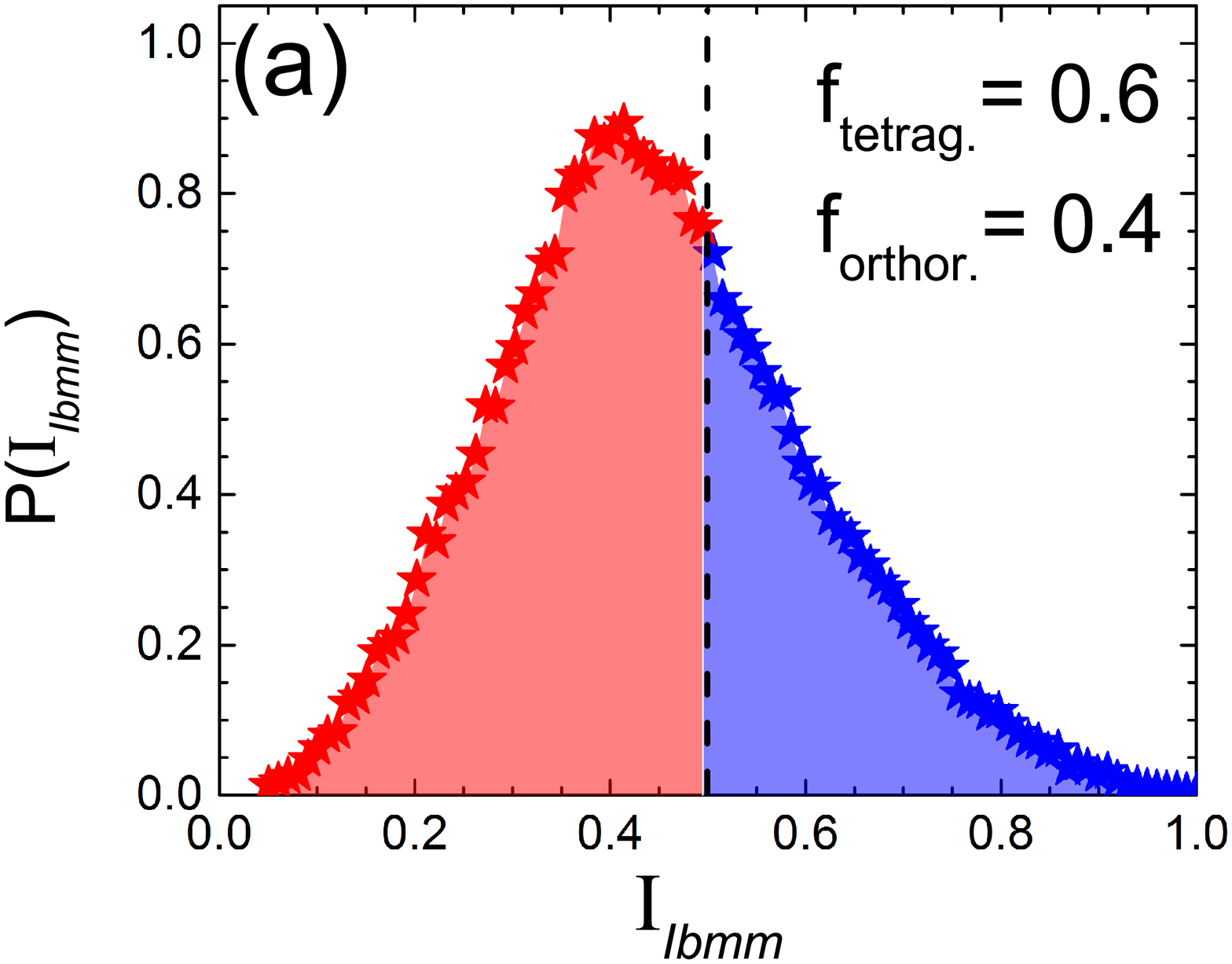} \hspace{-0.9cm}
\includegraphics[scale=0.24]{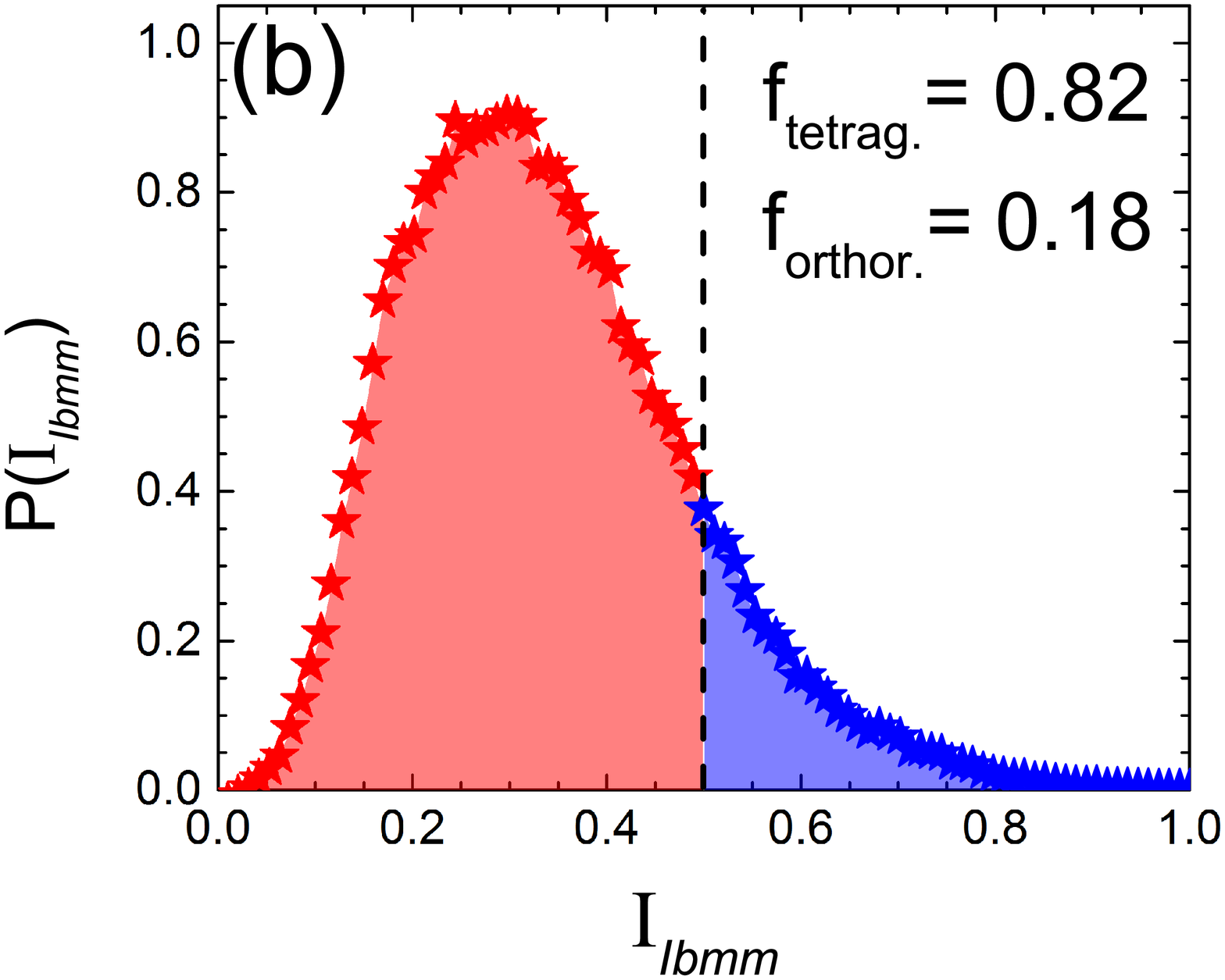}\hspace{-0.9cm}
\includegraphics[scale=0.24]{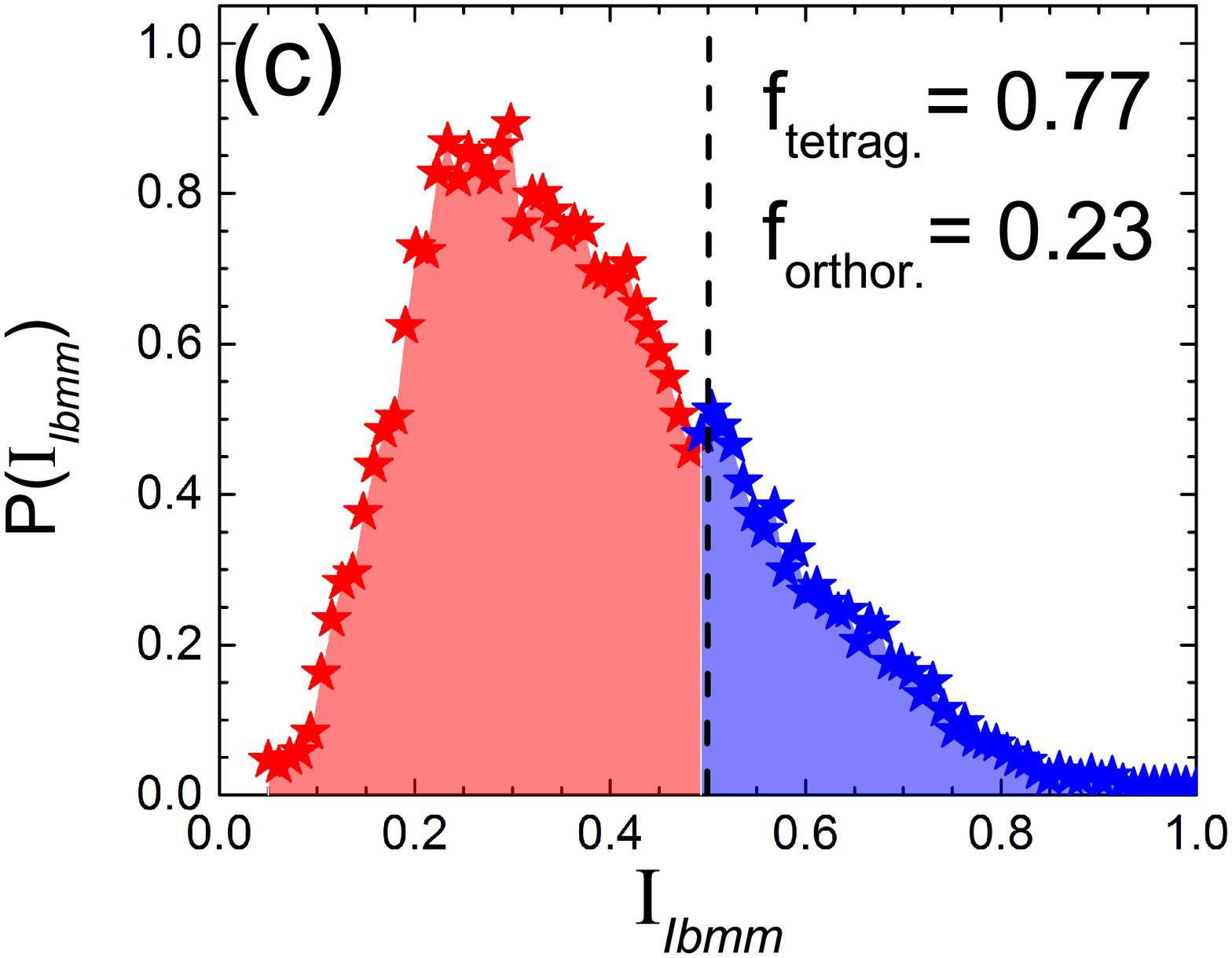} %\hspace{-0.7cm} 
%\vspace{-0.3cm}
\caption{(Color online) Average Probability distribution of orthorhombicity for samples with Bismuth concentration of {\bf (a)} $x=$0.18, {\bf (b)} $x=$0.24 and {\bf (c)} $x=$0.28. The blue and red shadowed regions separate the areas above and below half the intensity, from which the orthorhombic and tetragonal filling fractions can be estimated. }\label{fig_PorthoAv}
%\vspace{-0.4cm}
\end{figure*}

\begin{figure}[!htbp]
%\vspace{-0.7cm}
\hspace{-0.6cm}
\centering
\includegraphics[scale=0.34]{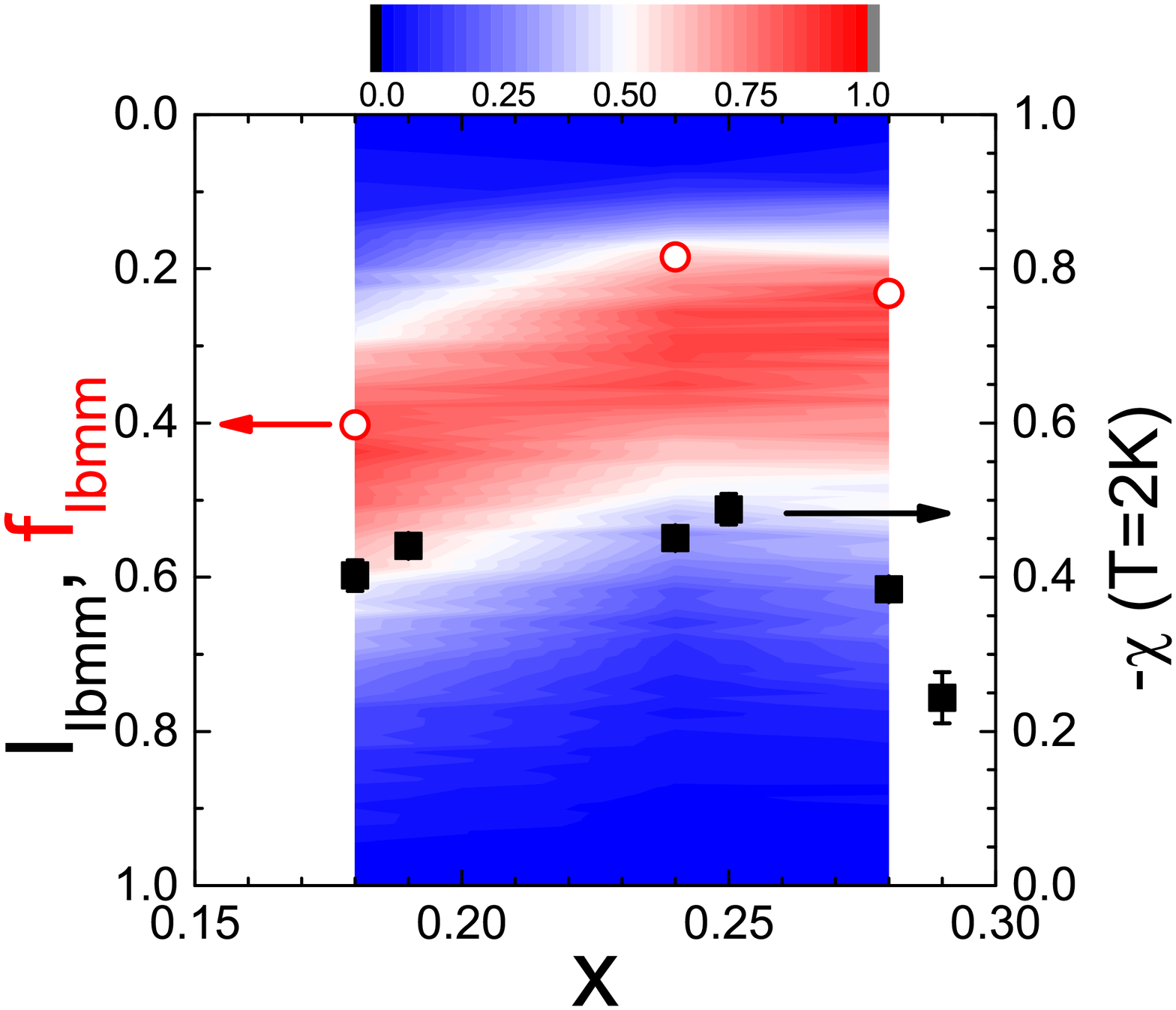} %\hspace{-0.7cm} 
\vspace{-0.3cm}
\caption{(Color online) The color scale represents the probability distribution of orthorhombicity, as a function of the orthorhombicity intensity (left scale) and Bi concentration (bottom scale). The red-open circles represent the orthorhombic filling fraction $f_{Ibmm}$ (left scale), calculated from integration of the probability distributions, with a 0.5 criteria. Black squares represent the superconducting volume fraction (right scale) determined from magnetic susceptibility measurements.} \label{figfrac}
%\vspace{-0.4cm}
\end{figure}

\section{Phase separation model}

The phase separation of tetragonal and orthorhombic polymorphs in BaPb$_{1-x}$Bi$_x$O$_3$ is presumably driven by changes in the relative Gibbs free energy of the two phases, both as a function of temperature and composition \cite{Ian1}. Such a scenario is illustrated schematically in fig. \ref{fig_spinodal}. The resulting morphology is reminiscent of spinodal decomposition, but the physical origin is somewhat different in this case, involving two competing phases. The point where the maximum tetragonal fraction is found, which for the case of BaPb$_{1-x}$Bi$_x$O$_3$ is coincident with optimal doping, $x_{opt}\approx$0.24, demarks a separatrix between a low-Bi orthorhombic phase, O(I), for compositions $x<x_{opt}$, and a Bi-rich orthorhombic phase, O(II), for compositions $x>x_{opt}$. It has been previously shown for various metallic precipitates embedded in metallic matrices (Cu-in-Al, Ag-in-Cu, Ag-in-Al, among others) that inhomogeneous strain can cause local variations in the free-energy, modifying phase equilibria \cite{lee1}. Therefore, it is reasonable to anticipate that the sharp distinctions in composition between the O(I) and O(II) phases will be blurred in practice (see panels (c) and (d) of figure \ref{fig_spinodal}). The resulting continuous variation in composition, and presumably lattice parameter, is consistent with results of recent x-ray and neutron diffraction measurements \cite{Cava1}. Consequently, the resistivity of the orthorhombic matrix in which the tetragonal polymorph resides, evolves continuously from that of a poor metal to a poor insulator.

%Previous x-ray and neutron diffraction studies have not provided indications of a scenario supporting compositional phase separation \cite{Cava1}. In these studies, the unit cell volume shows a monotonic variation as Bi composition is increased, suggesting that the material is homogeneously doped. However, within a phase separation model, the continuous variation of unit cell volume with Bi doping can be understood in terms of local variations of the Gibbs free energy. Since the free energy of each polymorph can be affected by strain \cite{lee1}, local variations in the local strain is anticipated to broaden or smear the otherwise sharp distinction in the variation of Bi composition. In addition, kinetics can affect phase separation in a solid, leading to local deviations from thermodynamic equilibrium.

%Althought this scenario would imply the observation The possible implication of such scenario are explain in the main manuscript.

\section{Dark-field TEM images}

In addition to the HRTEM images used in the analysis presented throughout this work, dark-field (DF) TEM images were also taken. These images also reveal the presence of broken-up stripes, with length scales consistent with the ones obtained through the $[110]_T/[101]_T$ filtered-and-reconstructed HRTEM images analysis described in the main text. Fig. \ref{fig_dark}(a,b) shows DF images for samples with Bi concentration of $x=$0.18 and 0.28, respectively, along the $[001]_T$ zone axis. These DF images were obtained using the $(110)$ reflection shown in the red circle in the insets to fig. \ref{fig_dark}(a,b). For both figures the size and distribution of the bright regions, imaged using reflection limited to the \textit{Ibmm} phase, mimics the small patchwork of coherent domains seen in the Fourier-filtered reconstruction. A stripy pattern, containing patches of 5-10nm, which for image in fig. \ref{fig_dark}(a) is more scattered throughout, is better seen in the image in fig. \ref{fig_dark}(b), and the length-scales observed are consistent with the ones found with the Fourier-filtered reconstruction. Note that only one of the four $(110)$ reflections are used in the DF image, while Fourier-filtered reconstruction uses all of the $[110]$ reflections in the reconstruction. This is most likely the cause of the $\left<1\bar{1}0\right>$ directionality seen in the image in figure \ref{fig_dark}(a).

%The size and distribution of the bright regions imaged using reflection limited to the \textit{Ibmm} phase mimics the small patchwork of coherent domains seen in the Fourier-filtered reconstruction. There are also larger patches of 5-10 nm scattered throughout. Note that only one of the four $(110)$ reflections are used in the DF image, while Fourier-filtered reconstruction uses all of the $[110]$ reflections in the reconstruction. This is most likely the cause of the $\left<1-10\right>$ directionality seen in the image in figure \ref{fig_dark}(a). 

\section{Intensity probability distributions}

We also computed the probability distribution of intensities of the $[101]_T/[110]_T$ filtered-and-reconstructed HRTEM images shown in Fig. 3 of the main text, and in Figs. \ref{fig_Gr2825}, \ref{fig_Gr2790} and \ref{fig_Gr3124} of this document. For each image, the intensity in each pixel, which might be labelled the intensity of ``orthorhombicity'', was normalized by the maximum intensity in the image, so that for all of the images this quantity goes from 0 to 1. Then, we computed the histogram of intensity, dividing the range of intensity into 100 sections. The vertical scale (counts) of each histogram is then normalized, so that for all the images this scale goes from 0 to 1. These curves are proportional to the probability distribution of intensity. The results for each image studied in this work are shown in fig. \ref{fig_Portho2825} for samples with Bi concentration $x=$0.18, in fig. \ref{fig_Portho2790} for samples with $x=$0.24, and in fig. \ref{fig_Portho3124} for samples with $x=$0.28. For each composition, all the normalized histograms were averaged, and the result of this averaging is shown in fig. \ref{fig_PorthoAv}. From these averaged histograms we estimated the orthorhombic filling fraction, $f_{Ibmm}$, as the normalized-integrated area above half the total intensity (as shown in the blue-shadowed areas in fig. \ref{fig_PorthoAv}).

The evolution of the orthorhombic volume fraction as a function of Bi concentration can be better visualized in fig. \ref{figfrac}. The left axis of this figure shows an inverted scale (from 1 to 0) of the orthorhombic intensity. The colors in the contour plot represent values of the probability of ``orthorhombicity'', i.e., the \textit{y} scale in the averaged histograms shown in fig. \ref{fig_PorthoAv}, as a function of orthorhombic intensity and Bi concentration. From this figure we can observe that the maximum of the probability distribution of ``orthorhombicity'' (red colors) is shifted toward lower values of orthorhombicity for optimal doping, compared to samples with lower and higher Bi concentrations. As a consequence, if we represent the orthorhombic volume fraction $f_{Ibmm}$ as the normalized-integrated area above half the total intensity, it is minimum at optimal doping, which means that the tetragonal fraction, $f_{I4/mcm}=1-f_{Ibmm}$ is maximum at this composition, with a value of 0.82$\pm$0.08. The evolution of the inverse of the orthorhombic fraction, ie., the tetragonal fraction, approximately tracks the evolution of the superconducting volume fraction, shown in the right scale and as black squares. These observations are consistent with results from x-ray and neutron diffraction experiments by E. Climent-Pascual \textit{et al.}\cite{Cava1} in polycrystalline samples, and suggest a direct connection between the tetragonal distortion and superconductivity.

%\section{Final comments}

%
% ****** End of file apstemplate.tex ******

%%%%%%%%%%%%%%%%%%%%%%%%%%%%%%%%%%%%%%%%%%%%%%%%%%%%%%%%%%

%\bibliography{1_BPBO_TEM_biblio}\vspace{-0.8cm}
%merlin.mbs apsrev4-1.bst 2010-07-25 4.21a (PWD, AO, DPC) hacked
%Control: key (0)
%Control: author (8) initials jnrlst
%Control: editor formatted (1) identically to author
%Control: production of article title (-1) disabled
%Control: page (0) single
%Control: year (1) truncated
%Control: production of eprint (0) enabled
\providecommand{\noopsort}[1]{}\providecommand{\singleletter}[1]{#1}%

\end{document}